\documentclass[fleqn,usenatbib]{mnras}

\usepackage{newtxtext,newtxmath}

\usepackage[T1]{fontenc}

\DeclareRobustCommand{\VAN}[3]{#2}
\let\VANthebibliography\thebibliography
\def\thebibliography{\DeclareRobustCommand{\VAN}[3]{##3}\VANthebibliography}

\usepackage{graphicx}	
\graphicspath{ {./Figures/} }
\usepackage{amsmath}	
\usepackage{multirow}
\usepackage{pdflscape}
\usepackage{longtable}
\usepackage{caption}
\usepackage{anyfontsize}    

\newcommand{\thisstar}{Gliese~12}
\newcommand{\thisstarb}{Gliese~12~b}

\defcitealias{GaiaDR3}{\textit{Gaia} DR3}

\title[Temperate Earth-size planet Gliese 12 b]{Gliese 12 b, A Temperate Earth-sized Planet at 12 Parsecs Discovered with \textit{TESS} and \textit{CHEOPS}}

\author[Dholakia, Palethorpe, et al.]{Shishir Dholakia,$^{1}$\thanks{These two authors contributed equally to this work \& should be considered joint first authors.}\thanks{E-mail: dholakia.shishir@gmail.com}
Larissa Palethorpe,$^{2,3,4}$\footnotemark[1]\thanks{E-mail: larissa.palethorpe@ed.ac.uk}
Alexander Venner,$^{1}$
Annelies Mortier,$^{5}$
\newauthor
Thomas G. Wilson,$^{6,7}$
Chelsea X. Huang,$^{1}$
Ken Rice,$^{2,3}$
Vincent Van Eylen,$^{4}$
Emma Nabbie,$^{1}$
\newauthor
Ryan Cloutier,$^{12}$
Walter Boschin,$^{8,9,10}$
David Ciardi,$^{11}$
Laetitia Delrez,$^{13,14,15}$
Georgina Dransfield,$^{5}$
\newauthor
Elsa Ducrot,$^{16}$
Zahra Essack,$^{17}$
Mark E. Everett,$^{18}$
Micha{\"e}l Gillon,$^{19}$
Matthew J. Hooton,$^{20}$
\newauthor
Michelle Kunimoto,$^{21,22}$
David W. Latham,$^{23}$
Mercedes L\'opez-Morales,$^{23}$
Bin Li,$^{24}$
Fan Li,$^{24}$
\newauthor
Scott McDermott,$^{25}$
Simon Murphy,$^{1}$
Catriona A. Murray,$^{26}$
Sara Seager,$^{21,27,28}$
Mathilde Timmermans,$^{13}$
\newauthor
Amaury Triaud,$^{5}$
Daisy A. Turner,$^{5}$
Joseph D. Twicken,$^{24}$
Andrew Vanderburg,$^{21}$
Su Wang,$^{24}$
\newauthor
Robert A. Wittenmyer,$^{1}$
and Duncan Wright$^{1}$
\\
(Affiliations can be found after the references)
}

\date{Accepted 2024 April 29. Received 2024 April 29; in original form 2024 January 30}

\pubyear{2024}

\begin{document}
\label{firstpage}
\pagerange{\pageref{firstpage}--\pageref{lastpage}}
\maketitle

\begin{abstract}

We report on the discovery of Gliese~12~b, the nearest transiting temperate, Earth-sized planet found to date. Gliese~12 is a bright ($V=12.6$ mag, $K=7.8$ mag) metal-poor M4V star only $12.162\pm0.005$~pc away from the Solar System with one of the lowest stellar activity levels known for an M-dwarf. A planet candidate was detected by \textit{TESS} based on only 3 transits in sectors 42, 43, and 57, with an ambiguity in the orbital period due to observational gaps. We performed follow-up transit observations with \textit{CHEOPS} and ground-based photometry with MINERVA-Australis, SPECULOOS, and Purple Mountain Observatory, as well as further \textit{TESS} observations in sector 70. We statistically validate Gliese~12~b as a planet with an orbital period of $12.76144\pm0.00006$ days and a radius of $1.0\pm{0.1}$ R$_\oplus$, resulting in an equilibrium temperature of $\sim$315K. Gliese~12~b has excellent future prospects for precise mass measurement, which may inform how planetary internal structure is affected by the stellar compositional environment. Gliese~12~b also represents one of the best targets to study whether Earth-like planets orbiting cool stars can retain their atmospheres, a crucial step to advance our understanding of habitability on Earth and across the Galaxy.

\end{abstract}




\section{Introduction}

Our knowledge of terrestrial planets akin to Earth around stars beyond our own has progressed vastly since the beginning of the 2010s. The \textit{Kepler} mission revealed that these small planets are abundant around low-mass M-dwarf stars \citep{Dressing2015, Muirhead2015}. This discovery has opened the doors to a wealth of study on small planets orbiting M dwarfs throughout the last decade. M dwarfs are especially well-suited for transiting planet detection and characterisation due to their frequency in our galaxy, small size, and low luminosities. These properties of M dwarfs result in advantages in detection efficiency due to nearer transiting systems, deeper transits, and higher \textit{a priori} geometric transit probability for temperate planets due to their closer orbital separations.

A key uncertainty in our current knowledge of these planets is whether they can retain their atmospheres, a prerequisite for their habitability \citep{Seager2010}. The recently-launched James Webb Space Telescope \citep[\textit{JWST};][]{JWST} is capable of detecting atmospheric features of such planets with transmission and emission spectroscopy, should they exist \citep{Rieke2015}. However, \textit{JWST} has so far observed featureless spectra for these planets \citep{Lustig-Yaeger2023, Lim2023, May2023, Zieba2023}. Though these null detections may intriguingly suggest that these planets lack atmospheres, their interpretation is complicated by signal-to-noise limitations and higher than anticipated effects of stellar contamination. Expanding the catalogue of terrestrial planets orbiting bright, nearby M dwarfs will be crucial for resolving this uncertainty.

The NASA Transiting Exoplanet Survey Satellite \citep[\textit{TESS};][]{Ricker2014} is performing an ongoing survey for transiting exoplanets across much of the sky. In contrast to the long-duration, deep-and-narrow observing strategy of \textit{Kepler}, \textit{TESS} observes a much larger 2300 deg$^2$ area of the sky using its four refractive 10 cm CCD cameras, with sky coverage divided in time into ``sectors" with durations of 27.4 days. This means that \textit{TESS} is sensitive to short-period planets orbiting bright stars. Furthermore, the red-optimised $600$-$1000$ nm \textit{TESS} bandpass increases its photometric sensitivity for cool stars such as M dwarfs \citep{Ricker2014}.

\textit{TESS} has discovered a significant number of small exoplanets orbiting M dwarfs throughout both its two-year prime mission and subsequent extended mission \citep[e.g.][]{Vanderspek2019, Luque2019, Gunther2019, Kostov2019, Winters2019, Cloutier2020, Cloutier2021, Shporer2020, Gan2020, Gilbert2020, Gilbert2023, Trifonov2021, Burt2021, Silverstein2022}. \textit{TESS} detections have demonstrated that terrestrial planets are common around the nearest M dwarfs \citep{Ment2023}, though the number of these planets detected around mid-late M dwarfs ($<$0.3~$R_\odot$) is lower than anticipated \citep{Brady2022}. Now in its sixth year of operations, \textit{TESS} continues to extend its observational baseline as it re-observes much of the sky. This enables the discovery of new exoplanets with longer orbital periods and, hence, lower equilibrium temperatures.

The CHaracterising ExOPlanets Satellite \citep[\textit{CHEOPS};][]{Benz2021} has found great success in taking precise transit photometry and determining accurate planetary parameters of \textit{TESS} discoveries that enables further characterisation by \textit{JWST} and other instruments \citep{Lacedelli2022,Tuson2023,Wilson2022, Fairnington2024, Palethorpe2024}. \textit{CHEOPS} utilises a 30 cm telescope to perform high precision, targeted photometry in order to refine transit ephemerides, ascertain orbital periods, and improve on the radius precision of planets detected by \textit{TESS} and other transit surveys \citep{Benz2021}, as well discovering new planets. The larger aperture of \textit{CHEOPS} compared to \textit{TESS} also provides high signal-to-noise transits that may be challenging or impossible to obtain from the ground. Due to \textit{TESS}'s scanning strategy, transits of longer period exoplanets are often missed, rendering the orbital period ambiguous. Follow-up observations by \textit{CHEOPS} can ascertain the orbital periods of these planets. A network of ground-based photometers and spectrographs also complements \textit{CHEOPS} in targeting \textit{TESS} Objects of Interest (TOIs) in order to confirm, validate, and refine the characteristics of the systems.

In this work we present the discovery of a temperate (F = $1.6\pm0.2$ S$_\oplus$) Earth-sized ($R_\text{b} =$ $1.0\pm{0.1}$ R$_\oplus$) planet orbiting the M-dwarf \thisstar{} found and validated through \textit{TESS} and \textit{CHEOPS} observations. \thisstar{} is bright ($V=12.6$~mag, $K=7.8$~mag), nearby ($12.162\pm0.005$~pc), and is characterised by low magnetic activity; these factors establish \thisstarb{} as a prime target for further characterisation, such as mass measurement via RV observations and atmospheric study through transmission spectroscopy.

We organise this work as follows: In Section \ref{sec:obs}, we detail the space- and ground-based photometry, spectroscopy, and imaging data for this target. In Section \ref{sec:star}, we compile and derive the properties of \thisstar{}. Section \ref{sec:validation} enumerates and precludes the possible scenarios of a false positive exoplanet detection with the available data, validating \thisstarb{} as a genuine exoplanet. We then describe our approach to modelling the transits of \thisstarb{} from the available data in Section \ref{sec:modelling}. In Section \ref{sec:discussion}, we comment on \thisstarb{}'s prospects for further study, including future mass measurement, atmospheric characterisation, and this target's utility to exoplanet demography. Finally, in Section \ref{sec:conclusions}, we summarise and conclude the work.

\section{Observations}
\label{sec:obs}

\subsection{Photometry}

\subsubsection{\textit{TESS}}
\label{sec:TESS_obs}

\thisstar{} (TIC 52005579, TOI-6251) was first observed by \textit{TESS} during ecliptic plane observations in sectors 42 and 43 of year 4 in its first extended mission \citep{Ricker2014, Ricker2021}, between 2021 August 21~-- 2021 October 11 (BJD~2459447.19~-- BJD~2459498.39). It was then re-observed during sector 57 of year 5 in the second \textit{TESS} extended mission between 2022 September 30~-- 2022 October 29 (BJD~2459852.85~-- BJD~2459881.62). Over a total of 82.2 days, \textit{TESS} observations of \thisstar{} were combined into 20s cadence target pixel frames\footnote{This star has been observed by \textit{TESS} as part of Guest Investigator programs G04033 (PI: Winters), G04148 (PI: Robertson), G04191 (PI: Burt), G04211 (PI: Marocco), and G04214 (PI: Cloutier) in year 4; programs G05087 (PI: Winters), G05109 (PI: Marocco), and G05152 (PI: Cloutier) in year 5; and program G06131 (PI: Shporer) in year 6.}. The data observed in sector 42 were taken on CCD 2 of camera 2, whilst the data observed in sectors 43 and 57 were taken on CCD 3 of camera 1.

The data were processed in the TESS Science Processing Operations Center (SPOC; \citet{Jenkins2016}) pipeline at NASA Ames Research Center. The SPOC conducted a transit search of the TESS data up to sector 57 on 2023 February 04 with an adaptive, noise-compensating matched filter \citep{Jenkins2002,Jenkins2010,Jenkins2020}. The search identified a Threshold Crossing Event (TCE) with 12.76148 d period. The TCE was reported with a Multiple Event Statistic (MES) S/N statistic of 8.0, above the lower limit of 7.1 required for a TCE to undergo further vetting. An initial limb-darkened transit model was fitted \citep{Li2019} and a suite of diagnostic tests were conducted to help assess the planetary nature of the signal \citep{Twicken2018}. The transit depth associated with the TCE was $1180\pm173$~ppm. Inspection of the SPOC Data Validation (DV) report reveals two possible periods for the transit signal, either 12.76~d or a factor of two larger (25.52~d), due to gaps in the photometry coincident with alternating transits.

\begin{figure*}
    \centering
    \includegraphics[width=\linewidth]{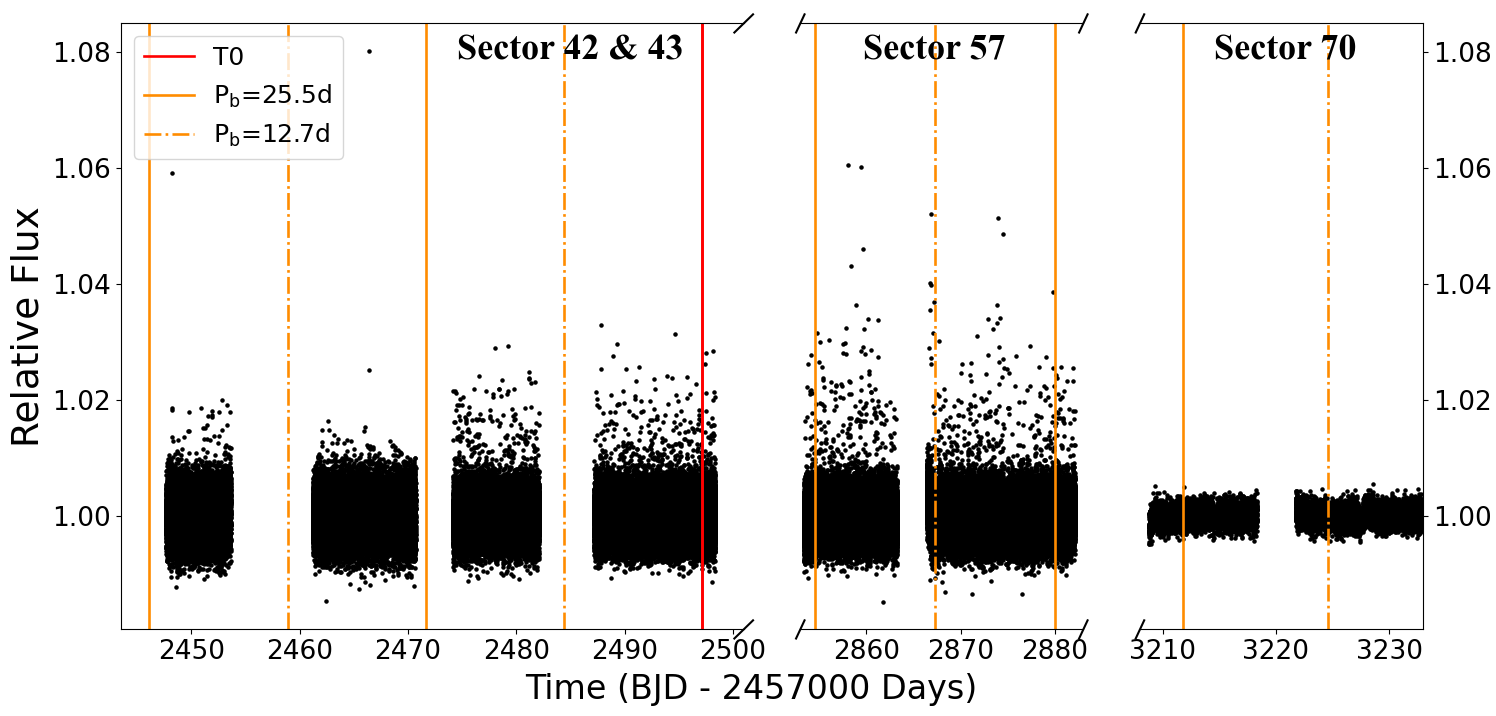}
    \caption{\textit{TESS} photometry of \thisstar{}. Figure shows the 20s cadence PDCSAP light curve for sectors 42, 43 \& 57 and the 120s cadence PDCSAP light curve for sector 70, where systematic errors have been removed, but the resulting light curve has not been corrected for low-frequency variations such as stellar activity. Over-plotted are the possible periods of the transiting planet, where the red solid line shows the transit centre, the orange dashed-dotted lines shows times consistent with a 12.76d and 25.52d period, and the solid orange line shows times consistent with only a 12.76d period.}
    \label{fig:TESS_LC}
\end{figure*}

\textit{TESS} photometry in sectors 42 and 43 was strongly affected by crossings of the Earth and Moon in the field of view \citep{TESS_DRN_S42, TESS_DRN_S43}. As a result observations of \thisstar{} in both sectors contain large mid-sector gaps, 7.6 d between BJD~2459453.64~-- BJD~2459461.24 in sector 42 and 5.1 d between BJD~2459482.11~-- BJD~2459487.19 in sector 43 (Figure \ref{fig:TESS_LC}), and there is likewise a large observing gap between the two sectors. The phasing of these gaps means that only one transit of the \thisstar{} planet candidate (BJD~2459497.18) was observed, with up to 3 transits falling outside of active observation. During sector 57 \textit{TESS} entered safe mode for 3.13 days between BJD~2459863.28~-- BJD~2459866.41; there are also three smaller gaps for data downlinks between BJD~2459860.58~-- BJD~2459860.80, BJD~2459867.25~-- BJD~2459867.47, and BJD~2459874.61~-- BJD~2459874.82, spanning a total of 0.63 d \citep{TESS_DRN_S57}. Two transits were successfully observed towards the beginning and end of sector 57 (BJD~2459854.51 and BJD~~2459880.03), however the location of a possible transit at BJD~2459867.27 falls into the data gap caused by the second data downlink.

The incidence of these data gaps, visualised in Figure \ref{fig:TESS_LC}, caused the initial ambiguity in the orbital period between 12.76~d or 25.52~d. Despite this uncertainty, the TCE passed an initial triage with the \textit{TESS}-ExoClass (TEC) classification algorithm\footnote{\url{https://github.com/christopherburke/TESS-ExoClass}} and was subsequently vetted by the \textit{TESS} team. The TCE was then duly promoted to \textit{TESS} Object of Interest (TOI; \citet{Guerrero2021}) planet candidate status and alerted as TOI-6251.01 on 2023 April 03. The TOI was alerted with the 25.52~d period, but a note was included on the factor-of-two orbital period ambiguity.

\textit{TESS} observed \thisstar{} again in sector 70, during its second ecliptic plane survey in year 6. Observations were captured with a 120s cadence on CCD 3 of camera 2 between 2023 September 20~-- 2023 October 16 (BJD~2460208.79 -- BJD~2460232.97). The \textit{TESS} sector 70 photometry covers two consecutive transits of TOI-6251.01 (also known as \thisstarb{}) separated by 12.76~d, confirming the short-period orbital solution.

For the photometric analysis, we download the \textit{TESS} photometry from the Mikulski Archive for Space Telescopes (MAST)\footnote{\url{https://archive.stsci.edu/tess}} and use the Presearch Data Conditioning Simple Aperture Photometry (PDCSAP; \citet{Stumpe2012, Stumpe2014, Smith2012}) light curve reduced by the SPOC. We use the quality flags provided by the SPOC pipeline to filter out poor quality data. 

\subsubsection{\textit{CHEOPS}}
\label{sec:CHEOPS_obs}

\begin{table*}
    \centering
    \renewcommand\multirowsetup{\raggedright}
    \caption{Log of \textit{CHEOPS} Observations of Gliese 12 b.  The column T\textsubscript{exp} gives the exposure time in terms of the integration time per image multiplied by the number of images stacked on-board prior to download. N\textsubscript{obs} is the number of frames. Effic. is the proportion of the time in which unobstructed observations of the target occurred. R\textsubscript{ap} is the aperture radius used for the photometric extraction. RMS is the standard deviation of the residuals from the best fit after the DRP has been applied. The variables in the final column are as follows: time, t; spacecraft roll angle, $\phi$, PSF centroid position, (x, y); smear correction, \texttt{smear}; aperture contamination, \texttt{contam}; image background level, \texttt{bg}.}
    \begin{tabular}{llllllllll}
         \hline\hline 
         Start Date & Duration & T\textsubscript{exp} & N\textsubscript{obs} & Effic. & File key & APER & R\textsubscript{ap} & RMS & Decorrelation \\ 
         (UTC) & (h) & (s) & & (\%) & & & (pixels) & (ppm) & \\
         \hline 
         2023-09-24T00:04 & 11.41 & 60 & 513 & 74.9 & CH\_PR240007 & DEFAULT & 25 & 801 & sin($\phi$), cos($\phi$), \\ & & & & & \_TG000101\_V0300 & & & & \texttt{contam, smear, bg} \\ 
         2023-10-06T18:20 & 12.76 & 60 & 481 & 62.8 & CH\_PR240007 & DEFAULT & 25 & 846 & t, t$^2$, x, x$^2$, y, y$^2$, sin($\phi$), \\ & & & & & \_TG000102\_V0300 & & & & sin($2\phi$), cos($\phi$), cos($2\phi$), \\ & & & & & & & & & \texttt{contam, smear, bg} \\
         2023-10-19T16:33 & 6.80 & 60 & 222 & 54.3 & CH\_PR240012 & DEFAULT & 25 & 752 & t$^2$, \texttt{contam, smear, bg} \\ & & & & & \_TG000701\_V0300 & & & & \\
         2023-11-14T06:30 & 6.92 & 60 & 241 & 57.9 & CH\_PR240012  & DEFAULT & 25 & 910 & t, x, y, sin($\phi$), cos($\phi$), \texttt{bg} \\ & & & & & \_TG000702\_V0300 & & & & \\
         2023-11-27T18:35 & 7.44 & 60 & 248 & 55.5 & CH\_PR240012 & DEFAULT & 25 & 1114 & t, x, y, sin($\phi$), cos($\phi$), \texttt{bg} \\ & & & & & \_TG000101\_V0300 & & & & \\
         \hline  
    \end{tabular}
    \label{tab:CHEOPS_obs}
\end{table*}

To confirm and characterise the planet in the Gliese 12 system we obtained five visits covering transits of planet b with the \textit{CHEOPS} spacecraft \citep{Benz2021} through the \textit{CHEOPS} AO-4 Guest Observers programmes with ID:07 (PI: Palethorpe) and ID:12 (PI: Venner). The observation log for our \textit{CHEOPS} observations is presented in Table~\ref{tab:CHEOPS_obs}.

The data were processed using the \textit{CHEOPS} Data Reduction Pipeline (DRP 14.1.3; \citealt{Hoyer2020}) that conducts frame calibration, instrumental and environmental correction, and aperture photometry using a pre-defined range of radii ($R$ = 15-40\arcsec). For all visits of Gliese 12 we selected the DEFAULT aperture, which has a radius of 25 pixels. The DRP produced flux contamination (see \citealt{Hoyer2020} and \citealt{Wilson2022} for computation and usage) that was subtracted from the light curves. We retrieved the data and corresponding instrumental bases vectors, assessed the quality using the \texttt{pycheops} Python package \citep{Maxted2022}, and decorrelated with the parameters suggested by this package, which can be found in Table \ref{tab:CHEOPS_obs}. Outliers were also trimmed from the light curves, with points that were 4$\sigma$ away from the median value removed. We then used these detrended data for further analysis. Table \ref{tab:photom_obs_period} shows which of the possible periods the various \textit{CHEOPS} transits cover, whilst three of the transits observed occurred when both a 12.76d and 25.5d period were possible, a further two transits were observed when only a 12.76d period was possible, confirming this as the orbital period of \thisstarb{}.

\begin{table*}
    \centering
    \renewcommand\multirowsetup{\raggedright}
    \caption{Log of ground-based photometric observations of \thisstarb{}.  The column T\textsubscript{exp} gives the exposure time in terms of the integration time per image multiplied by the number of images stacked on-board prior to download. N\textsubscript{obs} is the number of frames. The variables in the final column are as follows: time, t; position of the target star on the CCD (x, y), and the FWHM of the PSF.}
    \begin{tabular}{llllll}
         \hline\hline 
         Start Date & Duration & T\textsubscript{exp} & N\textsubscript{obs} & Observatory/Telescope & Decorrelation \\ 
         (UTC) & (h) & (s) & & & \\
         \hline 
         2023-09-11T11:39 & 3.852 & 30.0 & 197 & MINERVA T1 & t, x, y, airmass \\
         2023-09-11T11:39 & 3.852 & 60.0 & 266 & MINERVA T2 & t, x, y, airmass \\
         2023-09-24T01:30 & 7.019 & 10.0 & 1207 & SPECULOOS Europa & t$^2$, x$^2$ \\
         2023-11-01T13:01 & 3.000 & 55.0 & 210 & PMO & x, y, FWHM \\
         2023-11-27T00:01 & 4.251 & 10.0 & 754 & SPECULOOS Io & airmass, FWHM \\
         2023-11-27T00:02 & 4.234 & 10.0 & 741 & SPECULOOS Europa & t \\
         2023-11-27T00:02 & 4.239 & 10.0 & 751 & SPECULOOS Ganymede & t \\
         \hline  
    \end{tabular}
    \label{tab:ground_obs}
\end{table*}

\subsubsection{MINERVA-Australis}
\label{sec:MINERVA_obs}

We used three telescopes of the MINERVA-Australis array \citep{Addison2019} in clear bands to simultaneously observe a full transit of \thisstarb{} on the night of 2023 September 11.
Telescopes 1 and 2 are both equipped with ZWO1600 CMOS cameras, each with a field of view of 27' $\times$ 18', and plate scale of 0.67 arc-second per pixel. Telescope 3 is equipped with ZWO295 CMOS camera, with a field of view of 21' $\times$ 14', and a plate scale of 0.3 arc-second per pixel. We used exposure durations of 30 seconds for Telescope 1 and 2 and 60 seconds for telescope 3. We used \texttt{astroImageJ} \citep{Collins2017} to extract the light curves. The aperture radii are $\sim$8.7" for all three telescopes. The observation of telescope 3 was interrupted during the transit event, therefore did not achieve the necessary precision for the independent detection of the transit. Thus, we do not include data from telescope 3 for our global analysis. We detrend using the x and y positions, airmass and quadratic time series terms. A log of these observations and decorrelations can be found in Table \ref{tab:ground_obs}, and the possible period of the transiting planet covered by these observations can be found in Table \ref{tab:photom_obs_period}.

\subsubsection{SPECULOOS}
\label{sec:SPECULOOS_obs}

We observed two transits of Gliese 12 b with the SPECULOOS Southern Observatory \citep[SSO;][]{SPC_Jehin,SPC_Daniel, Gillon2018, Murray2020,SPC_Laeti}. SSO is composed of four 1m-class telescopes named after the Galilean moons (Io, Europa, Ganymede and Callisto), it is located at ESO Cerro Paranal Observatory in Chile. The telescopes are identical, each equipped with a 2K$\times$2K Andor CCD camera with a pixel scale of $0.35\arcsec$ per pixel, resulting in a field of view of $12\arcmin \times 12\arcmin$. We obtained a transit of Gliese 12 b on 2023 September 23 with Europa with 10s exposures in the \textit{Sloan-r'} filter, totalling 1213 measurements. A second transit was obtained simultaneously with Io, Europa and Ganymede on the 2023 November 26, also with 10s exposures in the \textit{Sloan-r'} filter. These observations represent 754, 741, and 751 measurements, respectively. All the data analysis was done using a custom pipeline built with the \texttt{prose}\footnote{\url{https://github.com/lgrcia/prose}} package \citep{2022_prose,prosesoft}. We performed differential photometry and the optimum apertures were $3.26\arcsec$, $3.45\arcsec$, $3.23\arcsec$, and $3.17\arcsec$ respectively for the Europa observation on the night of the 2023 September 23, and the Io, Europa, and Ganymede observations on the night of 2023 November 26. A log of these observations and their decorrelation parameters can be found in Table \ref{tab:ground_obs}, and the possible period of the transiting planet covered by these observations can be found in Table \ref{tab:photom_obs_period}.

\subsubsection{Purple Mountain Observatory}
\label{sec:PMO_obs}
We observed one transit with Purple Mountain Observatory (PMO) on the night of 2023 September 01. The telescope is an 80 cm azimuthal-mounting high precision telescope at Purple Mountain Observatory Yaoan Station in Yunnan Province, in the southwest of China. The telescope's field of view is 11.8 arcmin. It is a Ritchey-Chretien telescope with a 2048*2048 pixel PI CCD camera. The spatial resolution for each axis is 0.347 arcsec per pixel. We used the Ic broadband filter for the observation. The cadence of the observations was 55 seconds and the exposure time was 50 seconds. A log of these observations and their decorrelation parameters can be found in Table \ref{tab:ground_obs}, and the possible period of the transiting planet covered by these observations can be found in Table \ref{tab:photom_obs_period}.

\begin{table}
    \centering
    \renewcommand\multirowsetup{\raggedright}
    \caption{Log of photometric observations of \thisstarb{} relative to the possible period of the transiting planet. The Period column relates to the period ambiguity surrounding \thisstarb{}, where T$_0$ refers to the transit centre time of BJD 2459497.185, P$_1$ refers to the possible 12.76d period, and P$_2$ refers to the possible 25.5d period.}
    \begin{tabular}{llllll}
         \hline\hline 
         Date & Observatory/Telescope & Period \\ 
         (UTC) & & \\
         \hline 
         2023-09-11 & MINERVA (T1, T2) & T$_0 + 55$~P$_1$ \\
         2023-09-24 & CHEOPS & T$_0 + 56$~P$_1$ or T$_0 + 28$~P$_2$\\
         2023-09-24 & SPECULOOS (Europa) & T$_0 + 56$~P$_1$ or T$_0 + 28$~P$_2$ \\
         2023-10-06 & CHEOPS & T$_0 + 57$~P$_1$\\
         2023-10-19 & CHEOPS & T$_0 + 58$~P$_1$ or T$_0 + 29$~P$_2$\\
         2023-11-01 & PMO &  T$_0 + 59$~P$_1$\\
         2023-11-14 & CHEOPS & T$_0 + 60$~P$_1$ or T$_0 + 30$~P$_2$\\
         2023-11-27 & CHEOPS & T$_0 + 61$~P$_1$ \\
         2023-11-27 & SPECULOOS (Europa, & T$_0 + 61$~P$_1$ \\ & Ganymede, Io) & \\
         \hline  
    \end{tabular}
    \label{tab:photom_obs_period}
\end{table}

\subsection{Spectroscopy}

\subsubsection{TRES}
\label{sec:TRES_obs}

We obtained an additional 4 observations of \thisstar{} via the Tillinghast Reflector Echelle Spectrograph (TRES; \citet{Furesz2008}) on the 1.5m reflector at the Fred Lawrence Whipple Observatory in Arizona, USA. TRES is a fiber-fed echelle spectrograph with a spectral resolution of 44,000 over the wavelength range of 390–910nm. The observations were taken between 2016 October 13~-- 2021 September 11 as part of the M-dwarf survey of \citet{Winters2021}, using the standard observing procedure of obtaining a set of three science observations surrounded by ThAr calibration spectra. The science spectra were then combined to remove cosmic rays and wavelength calibrated using the ThAr spectra, with the extraction technique following procedures outlined in \citet{Buchhave2010}. The spectra had signal-to-noise ratios in the range 20-24 (average SNR = 21) at $\sim$716nm and are presented in Table \ref{tab:TRES_RVs}.

\subsubsection{HARPS-N}
\label{sec:HARPSN_obs}

We obtained 13 spectra for \thisstar{} with the HARPS-N spectrograph ($R=115000$) \citep{Cosentino_2012} installed on the 3.6m Telescopio Nazionale Galileo (TNG) at the Observatorio de los Muchachos in La Palma, Spain. These observations were taken between 2023 August 09~-- 2023 October 01 (BJD~2460165.66~-- BJD~2460218.56) as part of the HARPS-N Collaboration’s Guaranteed Time Observations (GTO) programme. The observational strategy consisted of exposure times of $1800$s per observation. We obtained spectra with signal-to-noise ratios in the range 12-35 (average SNR = 25) at 550nm.

The spectra were reduced with 2 different methods. The first used version 2.3.5 of the HARPS-N Data Reduction Software (DRS) \citep{Dumusque_2021}, with a M4 mask used in the cross-correlation function (CCF), resulting in an RV RMS of 2.60 ms\textsuperscript{-1} and RV precision of 2.95 ms\textsuperscript{-1}. The second method used was the line-by-line (LBL) method \citep{Artigau2022}. The \texttt{LBL} code \footnote{\url{https://github.com/njcuk9999/lbl}, version 0.61.0.} performs a simple telluric correction by fitting a TAPAS model \citep{Bertaux2014}, and has been shown to significantly improve the RV precision of M dwarfs observed with optical RV spectrographs \citep{cloutier2023masses}. The LBL method resulted in an average RV RMS of 2.69 ms\textsuperscript{-1}, which is similar to that of the CCF method, however the average RV precision significantly improved to 1.15 ms\textsuperscript{-1}. 

The HARPS-N data are presented in Table \ref{tab:RV_HARPSN}, which includes the radial velocities as well as the activity indicators: full width at half maximum (FWHM) of the CCF, the line Bisector Inverse Slope (BIS), H$\alpha$, and the log R$'_{\rm HK}$ converted from the S-index following \citet{SuarezMascareno2015}. RV observations with HARPS-N are still currently ongoing with the GTO programme and will be presented in a mass determination paper at a later date.

\subsection{High Resolution Imaging}
\label{sec:imaging_obs}

As part of our standard process for validating transiting exoplanets to assess the possible contamination of bound or unbound companions on the derived planetary radii \citep{ciardi2015}, we observed \thisstar{} with high-resolution near-infrared adaptive optics (AO) imaging at Keck Observatory and with optical speckle observations at WIYN.

\subsubsection{Optical Speckle at WIYN}
\label{sec:WIYN}

We observed \thisstar{} on 2018 November 21 using the NN-EXPLORE Exoplanet Stellar Speckle Imager (NESSI; \citet{Scott_2018}) at the WIYN 3.5~m telescope on Kitt Peak.  NESSI is a dual-channel instrument and obtains a simultaneous speckle measurement in two filters.  In this case, we used filters with central wavelengths of $\lambda_c = 562$ and 832~nm.  The speckle data consisted of 5 sets of 1000 40~ms exposures in each filter, centred on \thisstar{}.  These exposures were limited to a $256\times256$~pixel readout section in each camera, resulting in a $4.6\times4.6$~arcsecond field-of-view.  Our speckle measurements, however, are further confined to an outer radius of 1.2~arcseconds from the target star due the fact that speckle patterns lose sufficient correlation at wider separation.  To calibrate the shape of the PSF, similar speckle data were obtained of a nearby single star immediately prior to the observation of \thisstar{}.

\begin{figure}
    \centering
    \includegraphics[width=\linewidth]{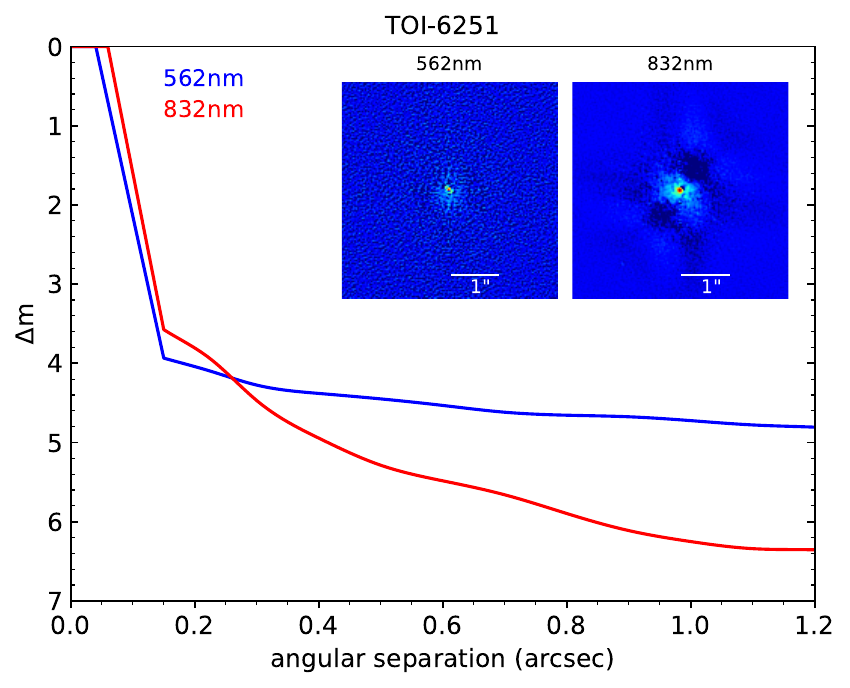}
    \caption{The 5$\sigma$ contrast curves of WIYN/NESSI speckle high-resolution images at 562\,nm (blue) and 832\,nm (red), with insets showing the central region of the images centred on \thisstar{}.}
    \label{fig:speckle}
\end{figure}

We reduced the speckle data using a pipeline described by \citet{Howell_2011}. Among the pipeline products is a reconstructed image of the field around \thisstar{} in each filter. The reconstructed images are the basis of the contrast curves that set detection limits on additional point sources that may lie in close proximity to \thisstar{}. These contrast curves are measured based on fluctuations in the noise-like background level as a function of separation from the target star.  The data, including reconstructed images, were inspected to find any detected companion stars to \thisstar{}, but none were found. We present the reconstructed images and contrast limits in Figure~\ref{fig:speckle}.

\subsubsection{Near-Infrared AO at Keck}
\label{sec:Keck}

\begin{figure}
    \centering
    \includegraphics[width=\linewidth]{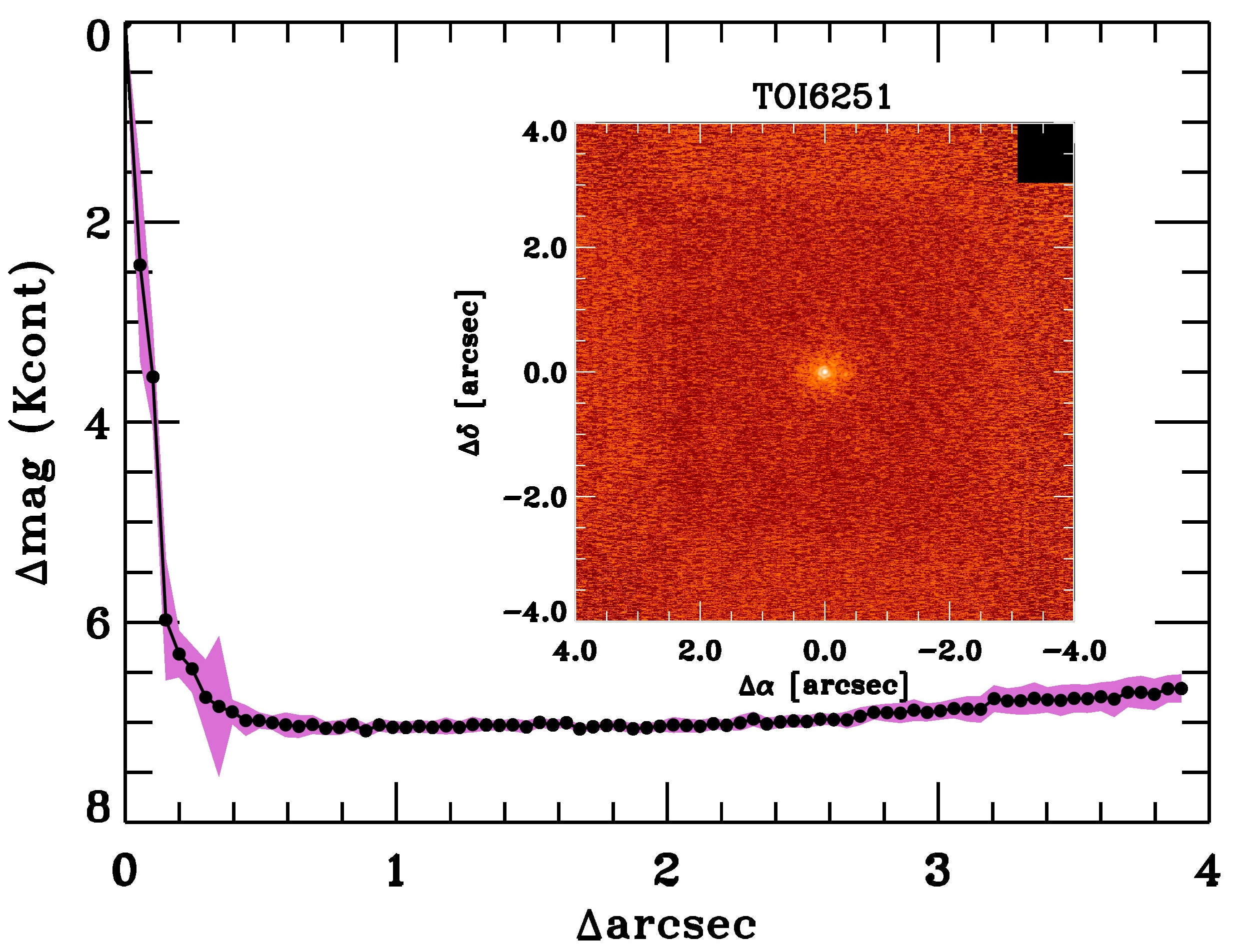}
    \caption{Companion sensitivity for the near-infrared adaptive optics imaging.  The black points represent the 5$\sigma$ limits and are separated in steps of 1 FWHM; the purple represents the azimuthal dispersion (1$\sigma$) of the contrast determinations (see text). The inset image is of the primary target showing no additional close-in companions.}\label{fig:ao_contrast}  
	\vspace{-0.5em}
\end{figure}

The observations were made with the NIRC2 instrument on Keck-II behind the natural guide star AO system \citep{wizinowich2000} on 2023 August 05 in the standard 3-point dither pattern that is used with NIRC2 to avoid the left lower quadrant of the detector which is typically noisier than the other three quadrants. The dither pattern step size was $3\arcsec$ and was repeated twice, with each dither offset from the previous dither by $0.5\arcsec$.  NIRC2 was used in the narrow-angle mode with a full field of view of $\sim10\arcsec$ and a pixel scale of approximately $0.0099442\arcsec$ per pixel.  The Keck observations were made in the $Kcont$ filter $(\lambda_o = 2.2706; \Delta\lambda = 0.0296~\mu$m) with an integration time in each filter of 1 seconds for a total of 9 seconds.  Flat fields were generated from a median average of dark subtracted dome flats. Sky frames were generated from the median average of the 9 dithered science frames; each science image was then sky-subtracted and flat-fielded.  The reduced science frames were combined into a single combined image using a intra-pixel interpolation that conserves flux, shifts the individual dithered frames by the appropriate fractional pixels; the final resolution of the combined dithers was determined from the full-width half-maximum of the point spread function; 0.049\arcsec.  To within the limits of the AO observations, no stellar companions were detected. The final $5\sigma $ limit at each separation was determined from the average of all of the determined limits at that separation and the uncertainty on the limit was set by the rms dispersion of the azimuthal slices at a given radial distance (Figure~\ref{fig:ao_contrast}).

\section{Stellar properties}
\label{sec:star}

\begin{table}
    \centering
    \caption{Stellar parameters of \thisstar{}.}
    \label{tab:star_params}
    \begin{tabular}{lll}
        \hline\hline
        Parameter & Value & Reference \\
        \hline
        Gliese & 12 & \citet{Gliese1956, Gliese1969} \\
        Giclas (G) & 32-5 & \citet{Giclas1959, Giclas1971} \\
        LHS & 1050 & \citet{Luyten1979} \\
        2MASS & 00154919+1333218 & \citet{2MASS} \\
        \textit{Gaia} DR3 & 2768048564768256512 & \citetalias{GaiaDR3} \\
        TIC & 52005579 & \citet{TIC} \\
        TOI & 6251 & \citet{Guerrero2021} \\
        \hline
        R.A. & 00:15:49.24 & \citetalias{GaiaDR3} \\
        Declination & +13:33:22.32 & \citetalias{GaiaDR3} \\
        $\varpi$ (mas) & $82.194\pm0.032$ & \citetalias{GaiaDR3} \\
        Distance (pc) & $12.162\pm0.005$ & Derived \\
        $\mu_\alpha$ (mas~yr$^{-1}$) & $+618.065\pm0.039$ & \citetalias{GaiaDR3} \\
        $\mu_\delta$ (mas~yr$^{-1}$) & $+329.446\pm0.034$ & \citetalias{GaiaDR3} \\
        RV (km~s$^{-1}$) & $+51.213\pm0.0036$ & \citet{Soubiran2018} \\
        $U$ (km~s$^{-1}$) & $-51.11\pm0.02$ & This work \\
        $V$ (km~s$^{-1}$) & $+27.31\pm0.03$ & This work \\
        $W$ (km~s$^{-1}$) & $-31.62\pm0.04$ & This work \\
        \hline
        Spectral Type & M3.5V & \citet{Lepine2013} \\
                      & M4V   & \citet{Newton2014} \\
        \hline
        $G$ (mag) & $11.399\pm0.003$ & \citetalias{GaiaDR3} \\
        $BP$ (mag) & $12.831\pm0.003$ & \citetalias{GaiaDR3} \\
        $RP$ (mag) & $10.227\pm0.004$ & \citetalias{GaiaDR3} \\
        $B$ (mag) & $14.265\pm0.038$ & \citet{TIC} \\
        $V$ (mag) & $12.600\pm0.042$ & \citet{TIC} \\
        $T$ (mag) & $10.177\pm0.007$ & \citet{TIC} \\
        $J$ (mag) & $8.619\pm0.002$ & \citet{2MASS} \\
        $H$ (mag) & $8.068\pm0.026$ & \citet{2MASS} \\
        $K_s$ (mag) & $7.807\pm0.020$ & \citet{2MASS} \\
        \hline
        $M$ ($M_\odot$) & $0.241\pm0.006$ & This work \\
        $R$ ($R_\odot$) & $0.269\pm0.008$ & This work \\
        $\rho$ (g~cm$^{-3}$) & $17.55^{+1.60}_{-1.42}$ & This work \\
        $\log g$ (cm~s$^{-2}$) & $4.96\pm0.03$ & This work \\
        $T_{\text{eff}}\, (K)$ & $3253\pm55$ & This work \\
        $L$ ($L_\odot$) & $0.0074\pm0.0008$ & This work \\
        $[\text{Fe/H}]$ (dex) & $-0.29\pm0.09$ & \citet{Maldonado2020} \\
        Age (Gyr) & $7.0^{+2.8}_{-2.2}$ & This work \\
        $\log R^{\prime}_{HK}$ & $-5.68\pm0.12$ & This work \\
        \hline
    \end{tabular}
\end{table}

Here we present our determinations of the stellar properties of \thisstar{}. Our adopted parameters are summarised in Table~\ref{tab:star_params}.

\subsection{Stellar parameters}

We estimate the mass and radius of \thisstar{} using the empirical relations of \citet{Mann2015, Mann2019}. To begin, we used the $K_s$-band apparent magnitude from 2MASS \citep[$7.807\pm0.020$~mag;][]{2MASS} and the stellar distance from the \textit{Gaia}~DR3 parallax \citep[$12.162\pm0.005$~pc;][]{GaiaDR3}, to derive the corresponding absolute $K_s$-band magnitude of $M_{Ks}=7.382\pm0.020$~mag. Using \citet[][equation~2 and tables~6,~7]{Mann2019} we convert this into a stellar mass $M=0.241\pm0.006~M_\odot$ (2.4\% uncertainty), and with \citet[][table~1 and equation~4]{Mann2015} we determine $R=0.269\pm0.008~R_\odot$ (3.0\% uncertainty). From these values for the mass and radius we derive the stellar density $\rho=17.55^{+1.60}_{-1.42}$~g~cm$^{-3}$ (9\% uncertainty) and surface gravity $\log g=4.96\pm0.03$~cm~s$^{-2}$.

Our stellar mass and radius agree well with the values adopted in the \textit{TESS} Input Catalogue \citep[$M=0.242\pm0.020~M_\odot$, $R=0.269\pm0.008~R_\odot$;][]{TIC}, which is unsurprising as the TIC Cool Dwarf list uses the \citet{Mann2015, Mann2019} $M_{Ks}$ relations to calculate these parameters \citep{TIC}. However, our estimate for the stellar density has a larger uncertainty than the TIC value ($17.46\pm0.12$~g~cm$^{-3}$). The constraint on this parameter (0.7\% uncertainty) appears implausibly precise considering the mass and radius uncertainties, so we prefer our characterisation of the stellar density precision. As we use priors on the stellar density in our transit fits this bears significance for our global models (on which see Section~\ref{sec:modelling}).

To further constrain the fundamental parameters of \thisstar{}, we perform a fit to the stellar spectral energy distribution (SED). We use the \texttt{astroARIADNE} Python package \citep{Vines2022} to fit the broadband photometry of \thisstar{} to the following stellar atmosphere models: BTSettl-AGSS2009 \citep{Allard2011}, Kurucz \citep{Kurucz1992}, and Castelli \& Kurucz (ATLAS9) \citep{Castelli2004}. This algorithm uses a Bayesian Model Averaging method to derive best-fit stellar parameters from the weighted average of each model's output. This is to mitigate the biases present in an individual model that correlate with the star's spectral type. Additionally, mass and age are estimated via MIST isochrones \citep{MISTModels}. We assign a broad Gaussian prior of $3376\pm157$~K on $T_{\text{eff}}$ from the TIC \citep{TIC}, and a prior of $\text{[Fe/H]}=-0.29\pm0.09$ from \citet{Maldonado2020}. From this analysis we find values of $T_{\text{eff}}=3253\pm55$~K and $L=0.0074\pm0.008~L_\odot$ for \thisstar{}, which we adopt in Table~\ref{tab:star_params}.

\subsection{Kinematics}

We calculate the space velocities of \thisstar{} relative to the Sun following \citet{Johnson1987}, adopting stellar positions and velocities as in Table~\ref{tab:star_params}. We find $(U,V,W)=(-51.11\pm0.02, +27.31\pm0.03, -31.62\pm0.04)$~km~s$^{-1}$, where $U$ is defined as positive in the direction of the galactic centre, $V$ is positive towards the galactic rotation, and $W$ is positive in the direction of the north galactic pole. Assuming values for the solar space velocities from \citet{Schonrich2010}, we then find space motions relative to the local standard of rest of $(U_{\text{LSR}},V_{\text{LSR}},W_{\text{LSR}})=(-40.0\pm0.8, +39.5\pm0.5, -24.3\pm0.4)$~km~s$^{-1}$. Following the membership probabilities established by \citet{Bensby2003} these kinematics, in particular the strongly positive value of $V_{\text{LSR}}$, are consistent only with thin disk membership for \thisstar{}. However, its kinematics are comparatively hot for a thin disk star; its $V_\text{tot}\equiv\sqrt{U_{\text{LSR}}^2+V_{\text{LSR}}^2+W_{\text{LSR}}^2}\approx61$~km~s$^{-1}$ belongs to the upper end of values found among nearby M dwarfs \citep{Newton2016, Medina2022}. This suggests a relatively old age for \thisstar{}.

We may extend this further by employing quantitative $UVW$-age relationships, such as that of \citet{AlmeidaFernandes2018} or \citet{Veyette2018}. These two relations are calibrated to the $UVW$ velocity dispersion-versus-isochrone age for Sun-like stars from the Geneva-Copenhagen survey \citep{Nordstrom2004, Casagrande2011}. For the measured $UVW$ space velocities of \thisstar{}, these relations return kinematic ages of $6.3^{+4.1}_{-3.1}$~Gyr \citep{AlmeidaFernandes2018} or $7.0^{+2.8}_{-2.2}$~Gyr \citep{Veyette2018}. These estimates are consistent with an older, most likely super-solar age for \thisstar{}. We adopt the age from the \citet{Veyette2018} relation for this star.

\subsection{Stellar activity}
\label{sec:stellaractivity}

We measured the S-index of \thisstar{} from the 13 HARPS-N spectra (Table~\ref{tab:RV_HARPSN}) and, following the method outlined in \citet{SuarezMascareno2015}, used them to compute the values of $\log R'_{HK}$. During the observation span of \thisstar{}, we find an average $\log R'_{HK}$ of $-5.68\pm0.12$. In comparison with the M-dwarf sample studied by \citet{Mignon2023}, \thisstar{} lies far towards the lower end of the observed distribution in $\log R'_{HK}$. The S-index and associated $\log R'_{\rm HK}$ are traditionally seen as an excellent indicator for a star’s magnetic cycle and overall activity level, implying a low level of stellar activity for \thisstar{}.

We also use the average value of $\log R'_{HK}$ to estimate the expected rotational period of \thisstar{}. Using the $\log R'_{HK}-P_{\text{rot}}$ relation for M3.5-M6 dwarfs from \citet{SuarezMascareno2018b}, we find an estimated rotational period of $132^{+37}_{-25}$~days. We point out that this estimate comes from a calibration relation and that the true rotation period may thus be outside these bounds. The rotational period of \thisstar{} has previously been measured from MEarth photometry as $78.5$~d \citep{Irwin2011} or $81.2$~d \citep{Newton2016}. There are various possibilities for the discrepancies between the \citet{SuarezMascareno2018b} prediction and measured values, for example, the manifestation of the rotation period in the data may not be the same as the physical rotation period \citep{Nava2020}. Alternatively, the $\log R'_{HK}$ of \thisstar{} varies over time due to a magnetic cycle \citep{Mignon2023}, therefore if it has been presently observed at a low value, then the predicted rotational period could be brought into agreement with the observed one.

Finally, we performed a Bayesian generalised Lomb-Scargle (BGLS) analysis \citep{Boisse2011, mortieretal15} on the \textit{TESS} SAP and PDCSAP photometry to search for any stellar variability, but did not detect any significant modulation. However, this analysis is adversely affected by the short $\sim$27d baseline of the \textit{TESS} sectors. Visual inspection of the \textit{TESS} lightcurves also shows no significant evidence for flares, which is expected for inactive M dwarfs in \textit{TESS} \citep{Medina2020} and consistent with a low magnetic activity level for \thisstar{}. This is consistent with the picture of an old stellar age \citep{Medina2022}.

\section{Planet validation}
\label{sec:validation}

\begin{figure}
    \centering
    \includegraphics[width=\linewidth]{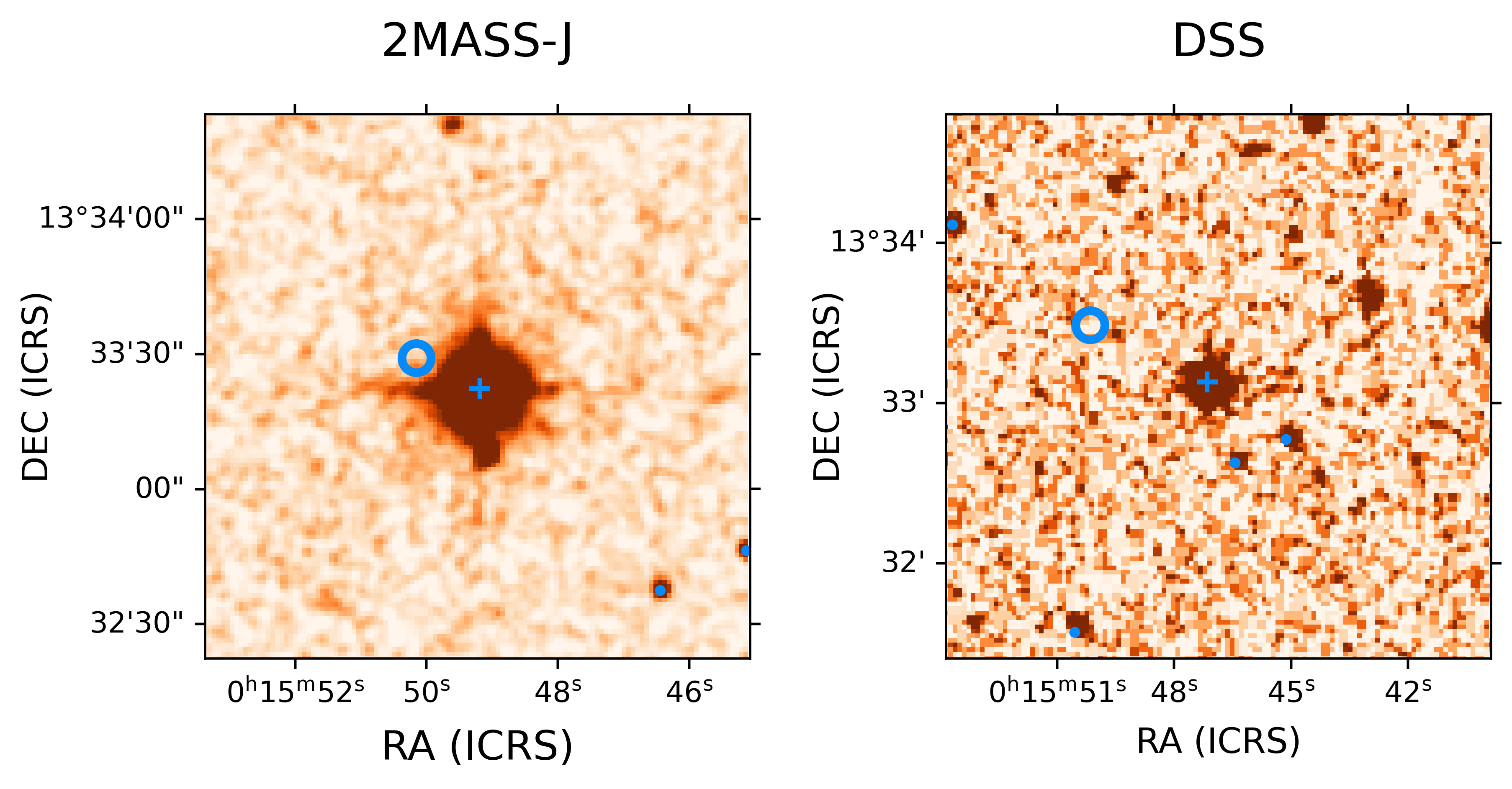}
    \caption{Historical images of the field surrounding \thisstar{}. \textbf{Left}: 2MASS J band image from 1998 \citep{2MASS} \textbf{Right}: DSS image from 1949 \citep{Lasker1990}. The blue circle indicate the location of \thisstar{} when it was observed by \textit{TESS}. The blue cross indicates the location of \thisstar{} when it was observed by 2MASS. Blue dots indicate all the other stars in the field of view within $\Delta$ T Mag 7.5 of \thisstar{}. A background eclipsing binary that can mimic the 1.1 ppt transit signal needs to be within $\Delta$ T Mag 7.5 of \thisstar{}. These images can be used to rule out the existence of such stars when \textit{TESS} is observing \thisstar{}. }
    \label{fig:finderimage}
\end{figure}

To ascertain the planetary nature of \thisstarb{}, we enumerate below several false positive scenarios that we can rule out from the available evidence:
\begin{enumerate}
    \item{\bf The discovery transit event is not an artifact of \textit{TESS} instrument systematics:} We can rule this out from five independent \textit{CHEOPS} visits and three ground-based observations that successfully detect a transit.\\
    \item{\bf \thisstarb{} is not a stellar companion:} The radial velocities from the TRES spectra, presented in Table \ref{tab:TRES_RVs}, vary by only 50 ms\textsuperscript{-1}, largely ruling out a stellar companion and motivating further observations with HARPS-N. From our HARPS-N radial velocity measurements outlined in Section \ref{sec:HARPSN_obs} and presented in Table \ref{tab:RV_HARPSN}, we find that the data has an RMS of only 2.60 m/s. From the peak-to-peak scatter in the RV observations, we can place an upper limit on the companion mass of $<10$ M$_\oplus$, well below the substellar limit. We also use \texttt{TRICERATOPS} \citep{Giacalone2021} to estimate the False Positive Probability (FPP) for \thisstarb{}. From \texttt{TRICERATOPS} we obtain a very low false positive probability of $< 5\times10^{-4}$.\\ 
    \item{\bf The transit event is not caused by a background eclipsing binary:} 
    We can preclude this scenario from archival imaging and high proper motion of \thisstar{}. In archival images of \thisstar{} taken by DSS and 2MASS, we see no resolved sources in the \textit{TESS} epoch sky position of \thisstar{} within $\Delta\text{T}<7.6$~mag out to 25 arcseconds (Figure \ref{fig:finderimage}). This is the minimum brightness difference required of an eclipsing binary to generate an eclipse that can be diluted to our observed transit depth of 1.1 ppt. We note that no sources exist in the Gaia Archive \citep{GaiaDR3} within 25 arcseconds down to G mag 20. Any resolvable sources within this radius would be too faint to cause significant contamination in the transit photometry. In addition, our high resolution AO (Section \ref{sec:imaging_obs}) reveals no companion sources within $\Delta$K mag of 6.5 farther than 0.5 arcsec from \thisstar{}. The SPOC data validation difference image centroid offset \citep{Twicken2018} locates the source of the transit signal within 9.9$\pm$8.5 arcsec of Gliese 12 and excludes all TIC objects within $\Delta$T mag of 7.6 as potential transit sources.\\
    
    \item{\bf \thisstarb{} does not host a widely separated companion that is an eclipsing binary:}
    If the flux from \thisstar{} were diluted by a bound hierarchical companion, the spectra in Section \ref{sec:HARPSN_obs} would reveal its presence unless it were smaller than \thisstar{}. Our False Positive Probability estimate with \texttt{TRICERATOPS} incorporates a computation of the probability of this scenario (PEB and PEBx2P) from the shape of the transit model. Our low FPP from \texttt{TRICERATOPS} therefore constrains this scenario as well. Furthermore, our high resolution adaptive optics and speckle imaging in Section \ref{sec:imaging_obs} reveal no companion sources down to K mag = 13.8 at 0.2 arcseconds. We can therefore confidently rule this scenario out.
\end{enumerate}

\section{Planet Modelling}
\label{sec:modelling}

\begin{table*}
    \caption{Planetary properties calculated from the combined transit fit. All limb darkening priors are normal for \texttt{emcee} and uniform for \texttt{dynesty}. We adopt the \texttt{emcee} values due to the more conservative parameter uncertainty on $R_\text{p}$.}
    \begin{tabular}{llll}
    \hline
    Parameter & \texttt{\textbf{emcee}} & \texttt{dynesty} & Source \\ 
    \hline
    \multicolumn{4}{l}{Fitted Parameters} \\
    \vspace{1mm}
    ~~~~$T_0$ (BJD) & $\mathbf{2459497.184 \pm{0.003}}$ &  $2459497.182_{-0.002}^{+0.003}$ & Measured (Uniform prior) \\
    \vspace{1mm}
    ~~~~$P$ (days) & $\mathbf{12.76144\pm{0.00006}}$ &  $12.76142_{-0.00006}^{+0.00005}$ & Measured (Uniform prior) \\
    \vspace{1mm}
    ~~~~$R_\text{p}/R_* $ & $\mathbf{0.034\pm{0.002}}$ &  $0.031\pm{0.001}$ & Measured (Uniform prior) \\
    \vspace{1mm}
    ~~~~$b$ & $\mathbf{0.80\pm{0.07}}$ & $0.67_{-0.05}^{+0.04}$ & Measured (Uniform prior) \\
    \vspace{1mm}
    ~~~~$\rho_{*}$ (g/cm$^3$) & $\mathbf{17.4\pm{0.6}}$ &  $17.4_{-1.4}^{+1.3}$ & Measured (Normal prior from Table \ref{tab:star_params}) \\
    \vspace{1mm}
    ~~~~$a/R_*$ & $\mathbf{53.1\pm{0.6}}$ &  $53.4_{-1.6}^{+1.5}$ & Measured (Uniform prior) \\
    \hline
    \multicolumn{4}{l}{Derived Properties} \\
    \vspace{1mm}
    ~~~~$R_\text{p}$ (R$_\oplus$) & $\mathbf{1.03\pm{0.11}}$ & $0.91\pm{0.04}$ & Derived \\
    \vspace{1mm}
    ~~~~$i$ (degrees) & $\mathbf{89.12\pm{0.07}}$ & $89.28_{-0.05}^{+0.06}$ & Derived \\ 
    \vspace{1mm}
    ~~~~$a$ (AU) & $\mathbf{0.066 \pm{0.002}}$ & $0.067\pm{0.003}$ & Derived \\
    \vspace{1mm}
    ~~~~$T_{\rm eq}$ (K) & $\mathbf{315\pm{5}}$ & $315\pm{10}$ & Derived (assuming A$_B$=0) \\
    \vspace{1mm}
    ~~~~$T_{\rm eq}$ (K) & $\mathbf{287\pm{5}}$ & $290\pm{9}$ & Derived (assuming A$_B$=A$_\oplus$) \\
    \vspace{1mm}
    ~~~~F (S$_\oplus$) & $\mathbf{1.6\pm{0.2}}$ & $1.6\pm{0.2}$ & Derived \\ 
    \hline
    \multicolumn{4}{l}{Limb Darkening Coefficients} \\
    \vspace{1mm}
    ~~~~q1$_{\rm TESS}$ & $\mathbf{0.40\pm{0.20}}$ & $0.60_{-0.32}^{+0.27}$ & Measured (\citet{Claret2017} prior) \\
    \vspace{1mm}
    ~~~~q2$_{\rm TESS}$ & $\mathbf{0.12\pm{0.07}}$ &  $0.32_{-0.23}^{+0.35}$ & Measured (\citet{Claret2017} prior) \\
    \vspace{1mm}
    ~~~~q1$_{\rm CHEOPS}$ & $\mathbf{0.60\pm{0.10}}$ &  $0.35_{-0.22}^{+0.27}$ & Measured (\citet{Claret2021} prior) \\
    \vspace{1mm}
    ~~~~q2$_{\rm CHEOPS}$ & $\mathbf{0.17\pm{0.06}}$ & $0.64_{-0.38}^{+0.26}$ & Measured (\citet{Claret2021} prior) \\
    \vspace{1mm}
    ~~~~q1$_{\rm MINERVA}$ & $\mathbf{0.60\pm{0.10}}$ & $0.60_{-0.35}^{+0.27}$ & Measured\\
    \vspace{1mm}
    ~~~~q2$_{\rm MINERVA}$ & $\mathbf{0.21\pm{0.06}}$ &  $0.53_{-0.35}^{+0.32}$ & Measured \\ 
    \vspace{1mm}
    ~~~~q1$_{\rm SPECULOOS}$ & $\mathbf{0.70\pm{0.20}}$ &  $0.71_{-0.33}^{+0.21}$ & Measured \\
    \vspace{1mm}
    ~~~~q2$_{\rm SPECULOOS}$ & $\mathbf{0.26\pm{0.07}}$ &  $0.61_{-0.37}^{+0.28}$ & Measured \\
    \vspace{1mm}
    ~~~~q1$_{\rm PMO}$ & $\mathbf{0.50\pm{0.20}}$ & $0.53_{-0.36}^{+0.32}$ & Measured\\
    \vspace{1mm}
    ~~~~q2$_{\rm PMO}$ & $\mathbf{0.26\pm{0.07}}$ & $0.51_{-0.34}^{+0.33}$ & Measured \\ 
    \hline
    \end{tabular}
    \label{tab:planet_results}
\end{table*}

We model the transits of \thisstarb{} using the available photometry from \textit{TESS}, \textit{CHEOPS}, MINERVA, and SPECULOOS. We first obtain our prior on the stellar density $\rho$ from Table~\ref{tab:star_params} to apply to our transit model.

Using the ensemble MCMC implemented in \texttt{emcee} \citep{emcee}, we simultaneously model the transit and apply least-squares detrending for the \textit{CHEOPS}, MINERVA, and SPECULOOS data. We use the transit model from \citet{Mandel2003}, implemented by \citet{kreidberg2015} in the BATMAN package, supersampling by a factor of 2 for the \textit{TESS} 20s cadence data, by a factor of 12 for the \textit{TESS} 120s cadence data, and by a factor of 4 for all \textit{CHEOPS} and ground-based data. We sample different quadratic limb darkening coefficients for the different bandpasses of the \textit{TESS}, \textit{CHEOPS}, MINERVA, SPECULOOS, and PMO photometry respectively. We obtain gaussian prior means and standard deviations on limb darkening coefficients from \citet{Claret2017, Claret2021}, using values from Table \ref{tab:star_params} for $T_{\text{eff}}$, $\log g$, and metallicity. For the \textit{TESS}, \textit{CHEOPS}, MINERVA, SPECULOOS, and PMO data we adopt \textit{TESS}, \textit{CHEOPS}, full optical, and Sloan r' filter bandpasses respectively in order to generate our prior.

We additionally used dynamic nested sampling through the \texttt{PyORBIT}\footnote{\url{https://github.com/LucaMalavolta/PyORBIT}, version 8.} package \citep{Malavolta2016, Malavolta2018}, which uses \texttt{dynesty} \citep{Speagle2020} to model planetary and stellar activity signals and obtain best-fit activity and planet parameters. \texttt{PyORBIT} makes use of the \texttt{BATMAN} python package for fitting the transit to the photometric data, where we assumed a quadratic stellar intensity profile for fitting the limb darkening coefficients, and an exposure time of 20.0\,s for the \textit{TESS} observations in sectors 42, 43, and 57, 120.0\,s for \textit{TESS} sector 70, 30.0\,s for the \textit{CHEOPS} observations, 30.0\,s and  60.0\,s for the MINERVA telescopes 1 and 2 observations respectively,  10.0\,s for the Europa, Ganymede, and Io SPECULOOS observations, and 55.0\,s for the PMO observations, as inputs to the light curve model. The priors placed on the limb-darkening coefficients were calculated using interpolation from \citet[][table~25]{Claret2017} for the \textit{TESS} light curves and from \citet[][table~8]{Claret2021} for the \textit{CHEOPS} light curves. For the \texttt{emcee} fit we apply Gaussian priors on limb darkening, while for the \texttt{dynesty} fit we apply these as Uniform priors. This is because recent empirical constraints have been in conflict with standard theoretical estimates of limb darkening for cool stars \citep{Patel2022}. We therefore use the \texttt{PyORBIT} model as confirmation that these potential problems do not significantly impact our posterior estimates.

\begin{figure*}
    \centering
    \includegraphics[width=0.45\linewidth]{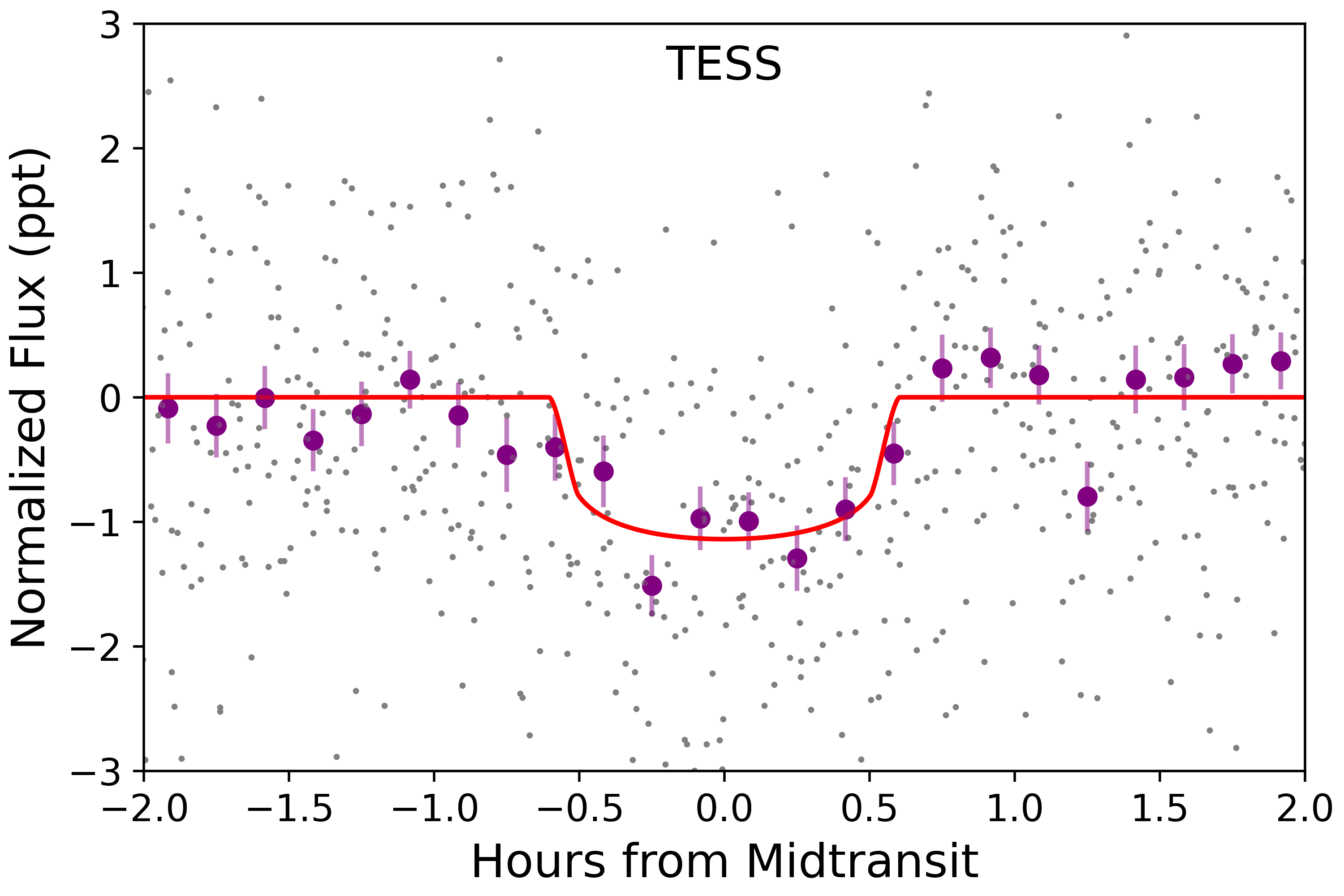}
    \includegraphics[width=0.45\linewidth]{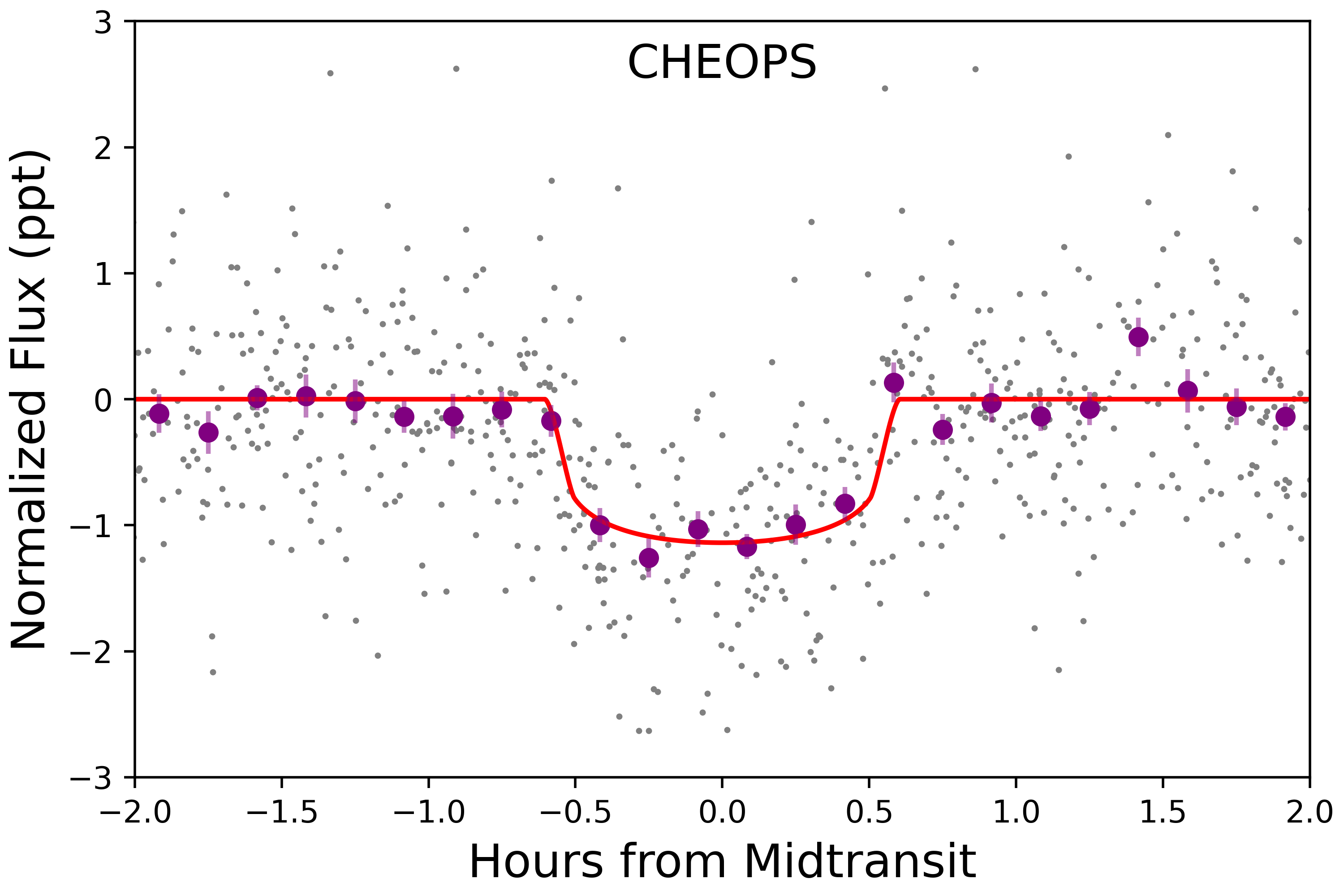}\\
    \includegraphics[width=0.33\linewidth]{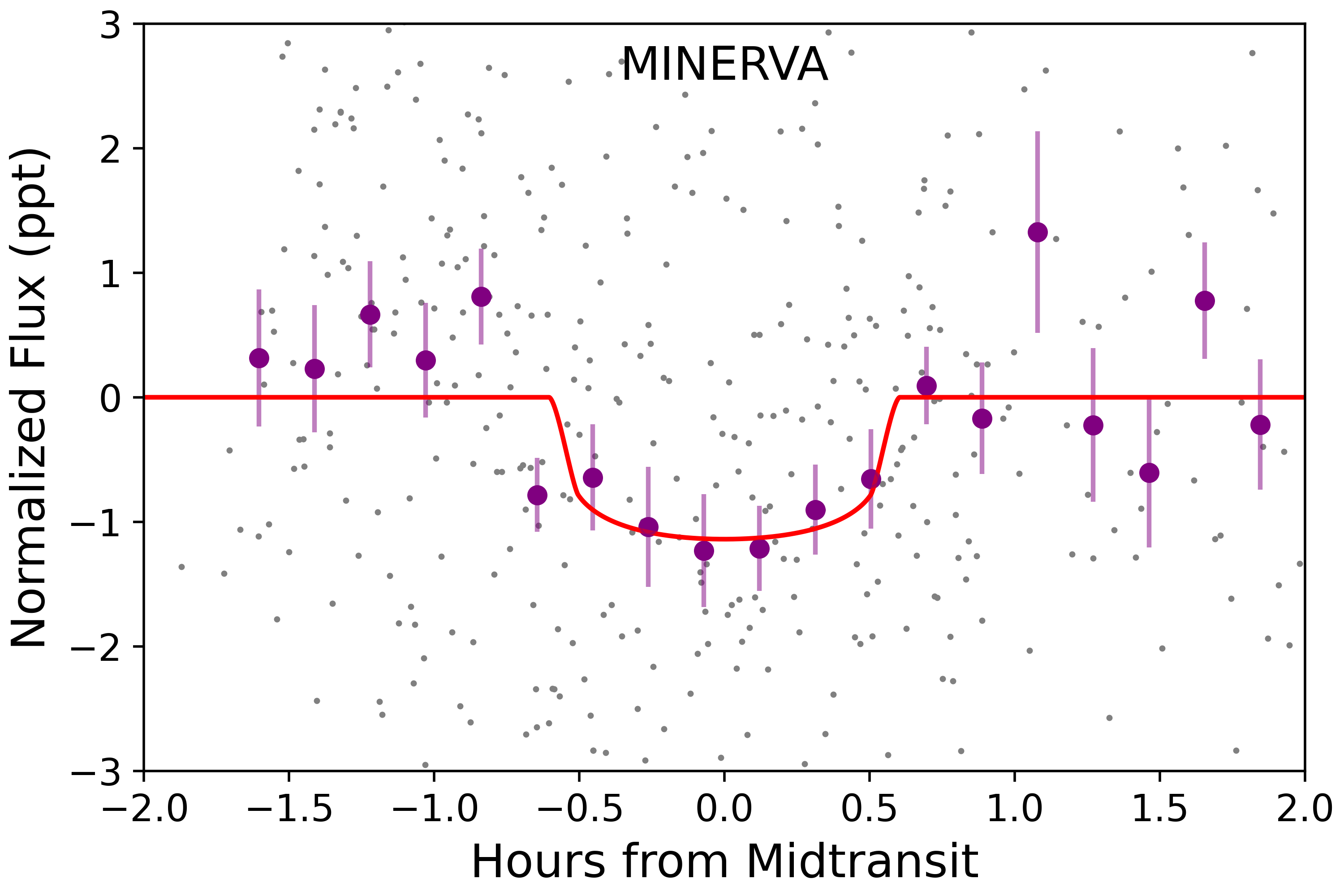}
    \includegraphics[width=0.33\linewidth]{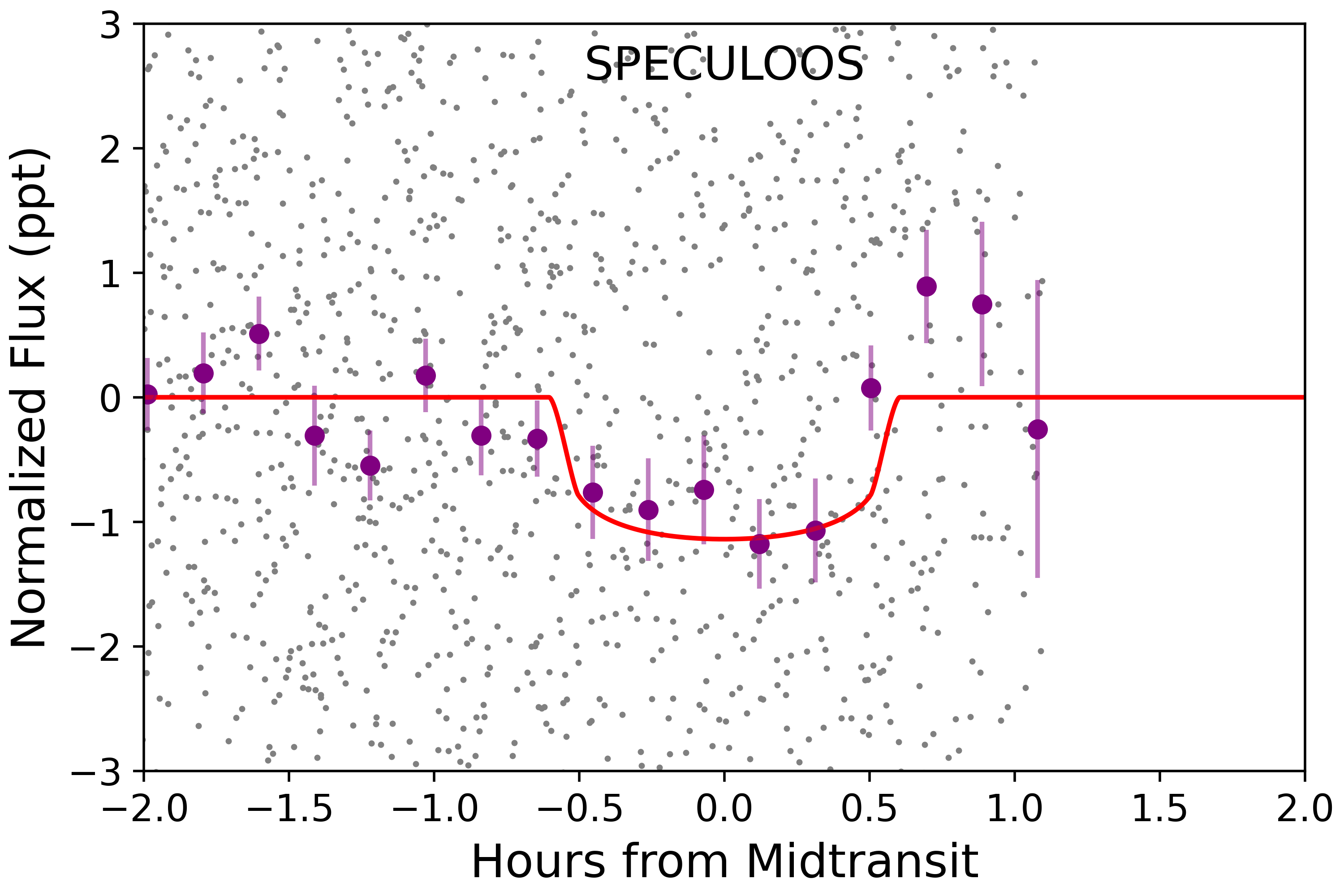}
    \includegraphics[width=0.33\linewidth]{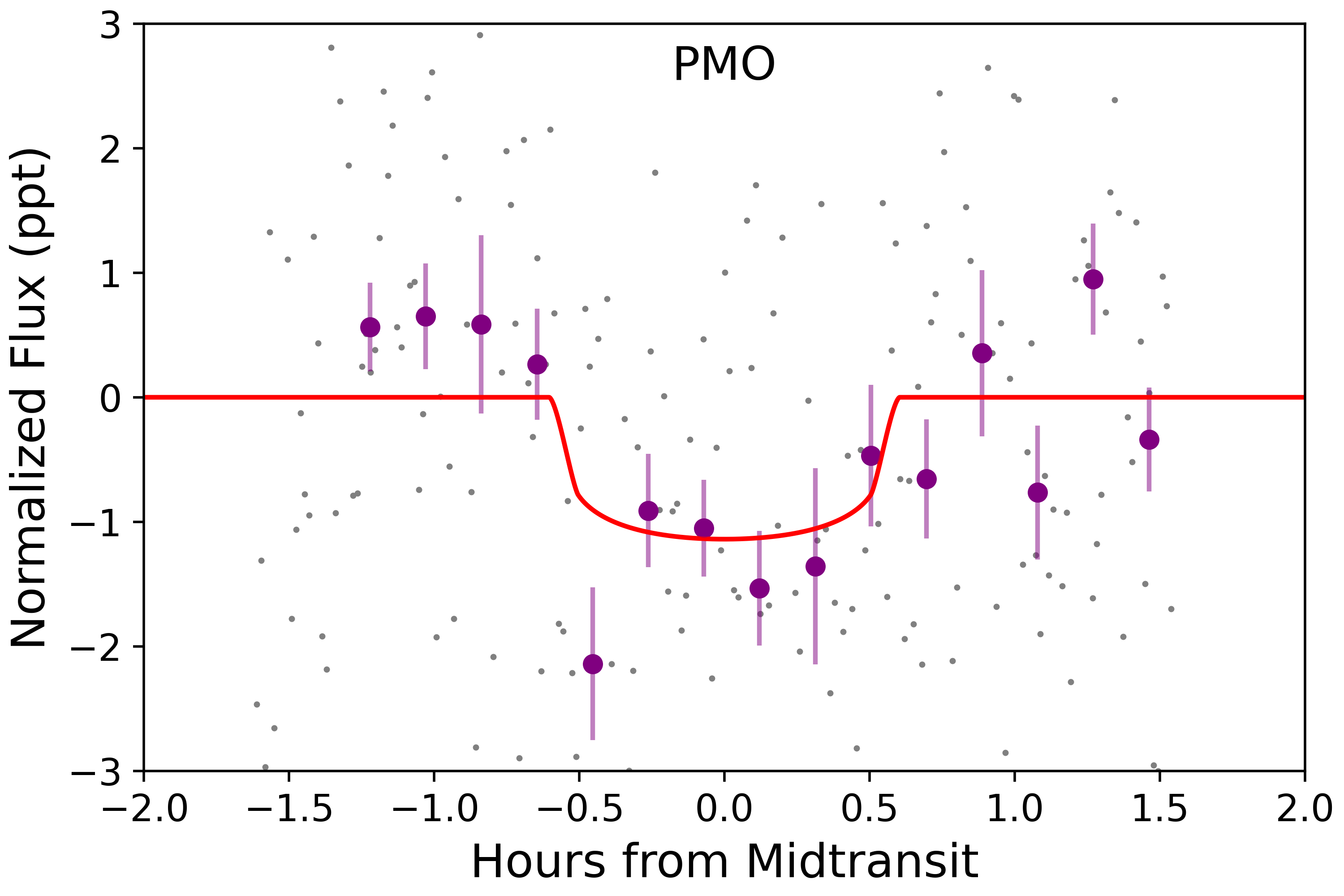}
    \caption{Phased transits of \thisstarb{} from fit results obtained with \texttt{emcee}. The red lines represent the transit from the best fit MCMC model. The purple data points are binned into 10 minute intervals for TESS and CHEOPS and 11.5 minute intervals for the ground-based data. The top left panel shows five phased \textit{TESS} transits, with the grey points representing the data binned down to 2 minute cadence. The top right panel shows five phased \textit{CHEOPS} transits of this target. The bottom left panel shows the MINERVA-Australis data. The bottom-center panel shows the SPECULOOS data, and the bottom right panel shows the PMO data.}
    \label{fig:transits}
\end{figure*}

Inferred planetary parameters from the fit for both \texttt{emcee} and \texttt{dynesty} are shown in Table \ref{tab:planet_results}, alongside the priors implemented. We find that the results of the two methods are consistent to 1$\sigma$ across all planetary parameters. We note that a difference in errors for our derived planetary radii is due to differences in our sampling techniques’ tendencies to handle deviations from Gaussianity in our posterior distributions. We adopt our final parameters from the model with the more conservative parameter uncertainty on $R_\text{p}$, which is the \texttt{emcee} fit. From this we find \thisstarb{} to be a small, temperate ($1.0 \pm{0.1}$ R$_\oplus$, $315 \pm{5}$ K) planet with a $12.76144\pm{0.00006}$ d orbital period. Figure \ref{fig:transits} shows the phase-folded light curves for the \textit{TESS}, \textit{CHEOPS}, MINERVA, SPECULOOS, and PMO data along with the best-fitting transit models using \texttt{emcee} sampling method. The phase-folded light curves and best-fitting transit models obtained from the \texttt{dynesty} sampling method for the data can be found in Appendix \ref{sec:add_figs} in Figure \ref{fig:dynesty_transits}.

\section{Discussion}
\label{sec:discussion}
\subsection{Prospects for mass measurement and interior composition}

We use the results of the global photometric analysis (see Table \ref{tab:planet_results}) to predict the expected planetary mass and radial velocity semi-amplitude of the planet. \citet{Otegi2020} present a mass-radius relation that is dependent on the density of the planet ($\rho$\textsubscript{b}):

\begin{equation}
    \mathrm{M_b [M_{\oplus}]}  = 
       (0.90 \pm 0.06)\mathrm{R_b} ^{(3.45 \pm 0.12)} \text{ if $\rho_\mathrm{b} > 3.3 \mathrm{gcm}^{-3}$}
\label{eq:mass}
\end{equation}

The high density case is applicable when the planet has a rocky composition, and assuming this, the planet would have a mass of $0.88_{-0.26}^{+0.39}$M$_\oplus$, leading to a semi-amplitude of only $0.57_{-0.17}^{+0.26}$ms$^{-1}$.

Given the brightness of the star, $V \rm{(mag)} = 12.600\pm0.042$, and following the early observations from HARPS-N, the predicted precision on the RV measurements with continued HARPS-N observations is 1.15\,ms$^{-1}$ with a 30min integration time. The planetary signal should thus be detectable with sufficient observations taken with a high-resolution high-stability spectrograph. Assuming a circular orbit and no correlated noise, we calculate that for a $3\sigma$ mass detection approximately 200 HARPS-N RV observations would be needed, with 30mins per observation, leading to a total of approximately 100 hours of telescope time. For a $5\sigma$ mass detection, approximately 600 RV observations would be needed, leading to 300hrs of telescope time. Whilst these calculations are specific to HARPS-N, there are other instruments capable of doing this, including those with better stability such as ESPRESSO.

The above estimates might be slightly optimistic since they ignore the possible impact of stellar activity. However, we highlight that thanks to the low activity level of \thisstar, the expected stellar-induced RV RMS should only be $0.27_{-0.14}^{+0.25}$ms\textsuperscript{-1} (using an extrapolation of the relations in \citet{Hojjatpanah2020}). This allows for the small planetary signal to be detectable in the data. Additional planets in the system, including non-transiting ones, may also be detectable through RV observations as multi-planet systems are common for M dwarfs, and compact multi-systems are also more common around low-metal stars \citep{Anderson2021}, such as \thisstar{}.

With a measured mass and thus bulk density, the planet's interior structure can be studied. The density and internal structures of Earth-like planets around metal-poor stars, such as \thisstarb{}, is important as recent observational studies have found potential compositional trends \citep{Wilson2022, Chen_2022} that may be imprints of planet formation or evolution \citep{owen2012planetary, Owen_2018}. \citet{Adibekayan2021} found that planet density is correlated with stellar iron mass fraction for Earth-like bodies, which provides evidence that the stellar compositional environment affects planetary internal structure. Different interior structure compositions due to formation environment variations could impact potential habitability \citep{Foley2016}. Planets around metal-poor stars may have small metallic cores and larger mantles compared to Earth and thus have weaker magnetic fields and increased volcanism. However, this is not well-understood as in the low-metallicity regime this trend is only anchored by two well-characterised planets \citep{Mortier2020, Lacedelli2022}, with none as small or cold as Earth. Therefore, by combining our precise radius measurement with an accurately measured mass of \thisstarb{}, we will be able to potentially verify a fundamental underlying process sculpting small planets, including our own Earth, across the Galaxy.

\subsection{Prospects for atmospheric characterisation}
\label{sec:atmo}

\begin{figure*}
    \centering
    \includegraphics[width=\linewidth]{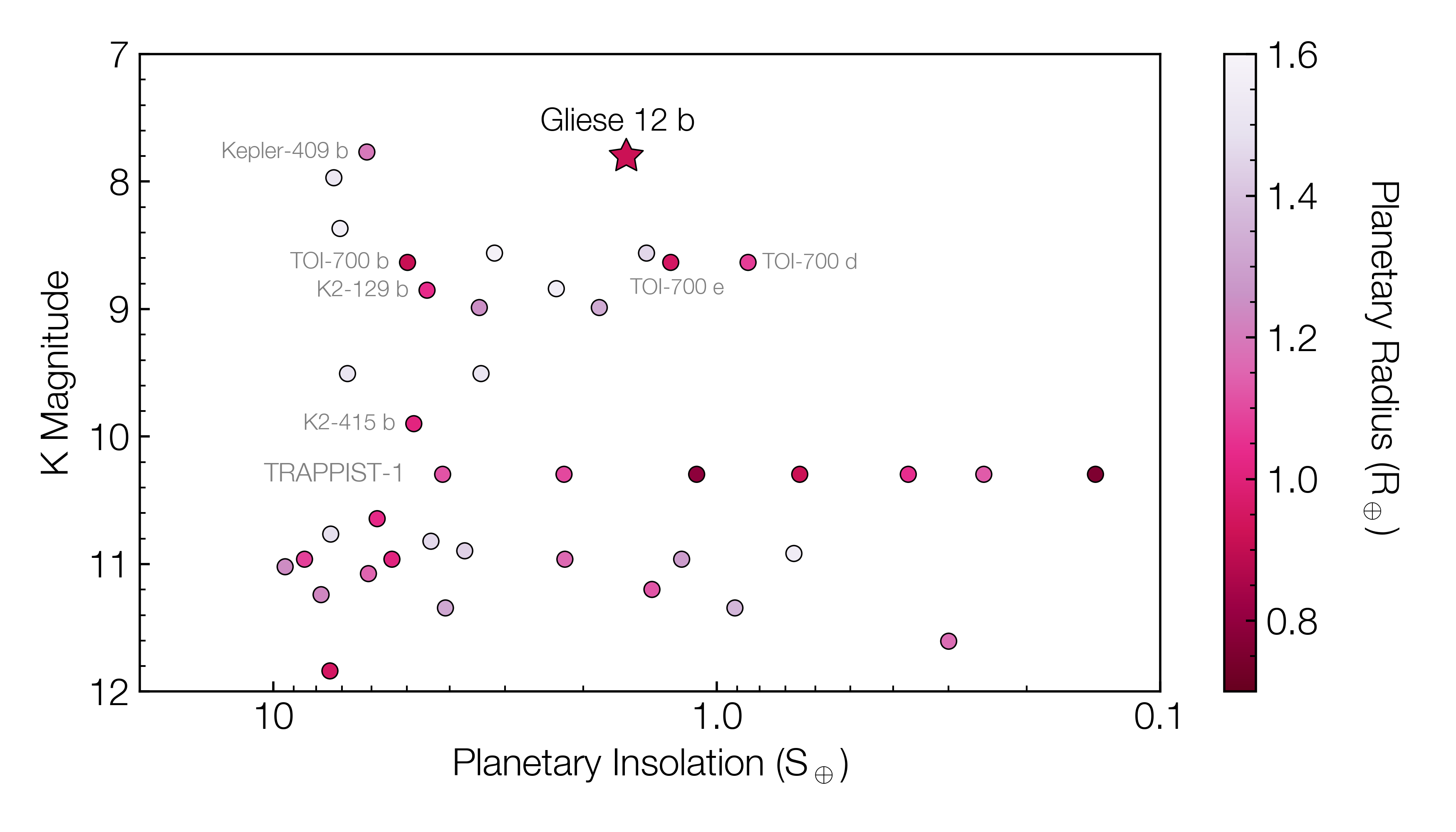}

    \caption{\thisstar{} is amongst the
brightest stars (in K band) hosting
a temperate Earth-sized planet.
Only planets with radii between
$0.7-1.6$ R$_\oplus$  and insolation flux
between $0.1$ and $10$ S$_\oplus$ on the Exoplanet Archive as of 03/01/2024 are shown
in the figure. The star represents \thisstarb{} and other notable Earth-sized planets are labeled. The TRAPPIST-1 planets in order from left to right are b, c, d, e, f, g, and h.}
    \label{fig:population}
\end{figure*}

\thisstarb{} is among the most amenable temperate ($1.6\pm{0.2}$ S$_\oplus$), terrestrial ($1.0\pm{0.1}$ R$_\oplus$) planets for atmospheric spectroscopy discovered to date. Its proximity to the Solar System ($12$ pc), high apparent brightness ($K=7.8$~mag), and relatively low activity level make it an ideal candidate for future transmission spectroscopy with \textit{JWST}. It is also unique in its parameter space for being exceptionally bright yet temperate (Figure \ref{fig:population}), making it ideal for characterisation.

So far, the TRAPPIST-1 system \citep{Gillon2017} and the TOI-700 system \citep{Gilbert2020,Rodriguez2020,Gilbert2023} have been the most well-studied exoplanet systems harbouring Earth-sized, temperate planets ($\lesssim 500$K, $\lesssim 1.6 R_{\oplus}$). The \textit{JWST} emission spectrum of TRAPPIST-1c disfavours a thick CO$_2$ dominated atmosphere, although the presence of higher molecular weight species cannot be constrained \citep{Zieba2023}. Other temperate planets in the TRAPPIST-1 system are planned to be targeted in \textit{JWST} transmission and emission spectroscopy. The Transmission Spectroscopy Metric (TSM; \citet{Kempton2018}) was created as a heuristic to assess the amenability of TESS planets to transmission spectroscopy measurements. The TSM value of \thisstarb{}, using the mass-radius relation presented in Equation \ref{eq:mass} is approximately 20. In comparison, the TRAPPIST-1 planets d, e, f, g and h present TSM values between 15 and 25. The TOI-700 system, however, has been shown to be just out of the reach of \textit{JWST}'s capabilities \citep{Suissa2020}, with a TSM of less than 5. \thisstar{} is both closer, brighter, and lower activity than TRAPPIST-1 or TOI-700, enabling uniquely constraining observations of its planet's atmosphere.

Just outside these equilibrium temperature and/or planetary radius bounds, \textit{JWST} transmission spectroscopy has returned mixed results for atmospheric detection on different planets around M dwarfs. LHS 475 b, though much hotter than \thisstarb{}, has been observed to have a featureless spectrum by \textit{JWST} so far \citep{Lustig-Yaeger2023}. For the hot super-Earth GJ 486b, the \textit{JWST} transmission spectrum contains tantalising features that may be interpreted either as a water rich atmosphere or contamination from unocculted starspots \citep{Moran2023}.

Contamination from stellar activity represents an important noise floor for detections of atmospheres around terrestrial planets, particularly for terrestrial planets around M dwarfs \citep{Rackham2018}. Most M dwarfs and many other stars present non-uniform  stellar photospheres due to the presence of large spots. These spots typically introduce photometric variability on a scale greater than 1\% of the total stellar flux. For these stars, unocculted spots introduce positive features in transmission spectra, which can mimic signatures due to molecular absorption in the planet's atmosphere \citep[e.g.][]{Moran2023}. This problem has been noted recently with \textit{JWST} in efforts to constrain temperate, terrestrial planet atmospheres for the TRAPPIST-1 system \citep{Lim2023}. \thisstar{} is well monitored by both ground based photometry and spectroscopy, showing a low activity level for M dwarfs and slow rotation in comparison to the \citet{Mignon2023} sample which featured only 7 stars with a lower $\log R'_{HK}$ (Section \ref{sec:stellaractivity}). Compared to the other similar systems, such as TOI-700 and TRAPPIST-1, the lack of visible starspot variability in the light curves, very low magnetic activity indicators, and slow rotation means that contamination in the planetary spectra from stellar spots will likely be far lower, making any interpretation of spectra features more tractable.

Whilst no high molecular weight atmosphere has yet been detected on such a planet, the atmospheres of Venus and the Earth in the Solar System hint at the possible diversity of such atmospheres elsewhere. M dwarfs are the most numerous stellar type in our galaxy, and terrestrial planets around them are commonplace \citep{Dressing2015}. The atmospheric compositions of true solar system analogues will be inaccessible for the foreseeable future. Therefore, studies of cool, rocky planets around M dwarfs -- similar in insolation and mass to Earth and Venus -- will provide insight into the formation of our Solar System and the diverse atmospheres of the terrestrial planets. H$_2$O, CO$_2$, and CH$_4$ features in the atmospheres of such planets would also be accessible to JWST \citep{Wunderlich2019}.

A key question in the study of the habitability of planets around M dwarfs is these planets' capacity to retain an atmosphere in the face of high stellar activity. Due to high XUV flux and flares from the stars, it has been hypothesised that the planetary atmospheres are photoevaporated efficiently, even for temperate planets \citep[e.g.][]{Lincowski2018,Zahnle2017}. The degree and extent to which this occurs is still poorly understood. \textit{JWST} atmospheric studies of \thisstarb{} therefore presents a valuable opportunity to make constraining measurements of a temperate, terrestrial planet's atmosphere, towards understanding the requisites of habitability for planets around M dwarfs.

\subsection{Prospects for Habitability}
\label{sec:habitable}
\begin{figure*}
    \centering
    \includegraphics[width=\linewidth]{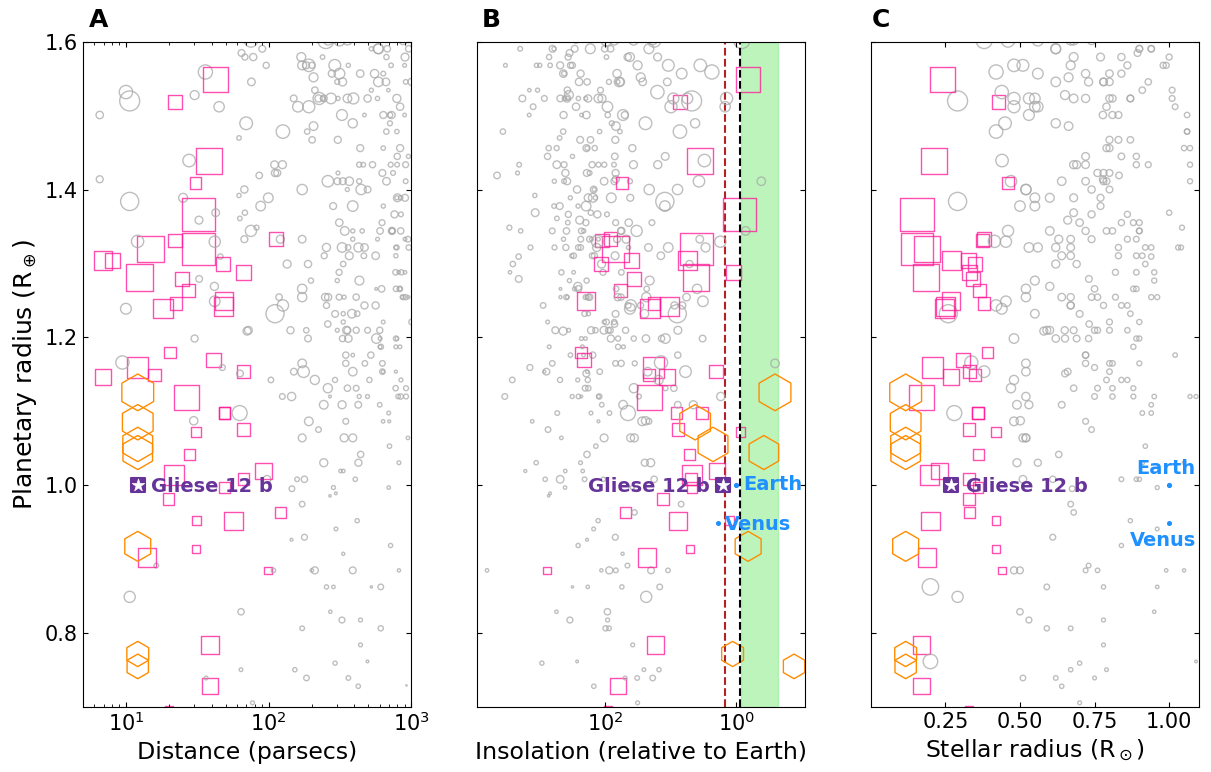}
    \caption{Planetary radius versus distance, insolation, and stellar radius, respectively. Planets were taken from the Extrasolar Planets Encyclopaedia, with those orbiting M dwarfs highlighted as pink squares, and the rest shown as grey circles. \thisstarb{} is highlighted by the white star in the purple square. The orange hexagons correspond to the TRAPPIST-1 planets; nearby planets around small stars that are most accessible to characterization by the James Webb Space Telescope (\textit{JWST}). The green shaded region in \textbf{B} is the M-dwarf habitable zone, the black dashed line corresponds to a runaway greenhouse atmosphere, and the dark red dashed line corresponds to recent Venus \citep{Kopparapu2013}. We note that this habitable zone is only appropriate for planets orbiting M dwarfs. The size of each marker is proportional to the transit depth and hence observational accessibility.}
    \label{fig:plot_HZ}
\end{figure*}

As the closest temperate, Earth-sized transiting planet to the Solar System, a key open question is the potential for \thisstarb{} to maintain temperatures suitable for liquid water to exist on its surface. With an insolation between the Earth and Venus's (F = $1.6\pm{0.2}$ S$_\oplus$), this planet's surface temperature would be highly dependant on its atmospheric conditions. It is also unknown whether \thisstarb{} is tide-locked or exists in a spin-orbit resonance, which may also impact the retention of water in the planet's evolutionary history \citep{Pierrehumbert2019}. \thisstarb{} occurs just inward of the habitable zone as defined by \citet{Kopparapu2013} due to the predicted efficiency of water-loss around M dwarfs. However, \thisstarb{} may well be within the Recent Venus limit of its host star (Figure \ref{fig:plot_HZ}) with an insolation flux of F = $1.6\pm{0.2}$ S$_\oplus$, less than 1$\sigma$ away from the 1.5 S$_\oplus$ limit calculated from \citet{Kopparapu2013}. This limit is regarded in that analysis as an optimistic habitable zone due to Venus's potential for habitability in the past history of the Solar system. Thus the available evidence does not rule out that \thisstarb{} is potentially habitable.

Due to its proximity to the Recent Venus limit, this target may be a valuable calibration point for theoretical estimates of water-loss, which define the inner edge of the habitable zone. In particular, \thisstarb{} is well suited to study the divergent evolutionary pathways of the atmospheres of Earth and Venus. While Earth retained its water, the runaway greenhouse effect on Venus led to water escape and photo-dissociation \citep{Ingersoll1969}. If a runaway greenhouse effect is in progress, signatures may be detectable in the UV band as an extended hydrogen exosphere \citep{Bourrier2017, 2Bourrier2017}. A null detection of water could hint that such a process has already occurred, as on Venus. On the other hand, a detection of water would imply that the water-loss boundary is inward of the estimates by \citet{Kopparapu2013}.
 
\section{Conclusions}
\label{sec:conclusions}

In this paper, we have reported on the discovery and validation of a temperate Earth-sized planet transiting the nearby M dwarf \thisstar{}. During \textit{TESS} sectors 43, 57, only 3 transits of the planet were observed, with no transits observed in sector 42. This allowed for a period of 12.76 d or 25.5 d, but with the further addition of \textit{TESS} sector 70 and \textit{CHEOPS} follow-up we were able to determine the actual period. Combining these transits with further ground-based photometry from MINERVA, SPECULOOS, and PMO allowed us to determine precise planetary parameters for \thisstarb{}, indicating that \thisstarb{} is a $1.0\pm{0.1}$ R$_\oplus$ planet with a $12.76144\pm{0.00006}$ day period, and an effective temperature lower than that of the majority of known exoplanets ($\sim$315K). 

\thisstarb{} is therefore a prime target for future detailed characterisation studies. Its prospects for a precise mass measurement are reasonable and the star has some of the lowest activity levels amongst known M dwarfs. It is also a unique candidate for both atmospheric and stellar study. Further analysis of the \thisstar{} system will allow us to understand evolutionary and compositional trends, which is important as we try to infer the number of true-Earth analogues on our journey to understanding our own place in the Universe.

\section*{Acknowledgements}

CXH and AV are supported by ARC DECRA project DE200101840. VVE is supported by UK's Science \& Technology Facilities Council (STFC) through the STFC grants ST/W001136/1 and ST/S000216/1.

This paper includes data collected by the \textit{TESS} mission, which are publicly available from the Mikulski Archive for Space Telescopes (MAST). Funding for the \textit{TESS} mission is provided by NASA’s Science Mission directorate. We acknowledge the use of public \textit{TESS} Alert data from pipelines at the \textit{TESS} Science Office and at the \textit{TESS} Science Processing Operations Center. Resources supporting this work were provided by the NASA High-End Computing (HEC) Program through the NASA Advanced Supercomputing (NAS) Division at Ames Research Center for the production of the SPOC data products. 

This work makes use of data from the European Space Agency (ESA) \textit{CHEOPS} mission, acquired through the \textit{CHEOPS} AO-4 Guest Observers Programmes ID:07 (PI: Palethorpe) and ID:12 (PI: Venner). \textit{CHEOPS} is an ESA mission in partnership with Switzerland with important contributions to the payload and the ground segment from Austria, Belgium, France, Germany, Hungary, Italy, Portugal, Spain, Sweden and the United Kingdom. We thank support from the \textit{CHEOPS} GO Programme and Science Operations Centre for help in the preparation and analysis of the \textit{CHEOPS} observations. This research has made use of the Exoplanet Follow-up Observation Program (ExoFOP) website, which is operated by the California Institute of Technology, under contract with the National Aeronautics and Space Administration under the Exoplanet Exploration Program. 

This work makes use of data from MINERVA-Australis. MINERVA-Australis is supported by Australian Research Council LIEF Grant LE160100001, Discovery Grants DP180100972 and DP220100365, Mount Cuba Astronomical Foundation, and institutional partners University of Southern Queensland, UNSW Sydney, MIT, Nanjing University, George Mason University, University of Louisville, University of California Riverside, University of Florida, and The University of Texas at Austin. We acknowledge and pay respect to Australia’s Aboriginal and Torres Strait Islander peoples, who are the traditional custodians of the lands, waterways and skies all across Australia. In particular, we pay our deepest respects to all Elders, ancestors and descendants of the Giabal, Jarowair, and Yuggera nations, upon whose lands the MINERVA-Australis facility is situated.

We thank Ye Yuan and Jian Chen for their assistance in obtaining and reducing the Purple Mountain Observatory photometry data.

Based on data collected by the SPECULOOS-South Observatory at the ESO Paranal Observatory in Chile.The ULiege's contribution to SPECULOOS has received funding from the European Research Council under the European Union's Seventh Framework Programme (FP/2007-2013) (grant Agreement n$^\circ$ 336480/SPECULOOS), from the Balzan Prize and Francqui Foundations, from the Belgian Scientific Research Foundation (F.R.S.-FNRS; grant n$^\circ$ T.0109.20), from the University of Liege, and from the ARC grant for Concerted Research Actions financed by the Wallonia-Brussels Federation. This work is supported by a grant from the Simons Foundation (PI Queloz, grant number 327127).
This research is in part funded by the European Union's Horizon 2020 research and innovation programme (grants agreements n$^{\circ}$ 803193/BEBOP), and from the Science and Technology Facilities Council (STFC; grant n$^\circ$ ST/S00193X/1, and ST/W000385/1).
The material is based upon work supported by NASA under award number 80GSFC21M0002. This publication benefits from the support of the French Community of Belgium in the context of the FRIA Doctoral Grant awarded to M.T. This publication benefits from the support of the French Community of Belgium in the context of the FRIA Doctoral Grant awarded to M.T.

Some of the data presented herein were obtained at Keck Observatory, which is a private 501(c)3 non-profit organization operated as a scientific partnership among the California Institute of Technology, the University of California, and the National Aeronautics and Space Administration. The Observatory was made possible by the generous financial support of the W. M. Keck Foundation. The authors wish to recognize and acknowledge the very significant cultural role and reverence that the summit of Maunakea has always had within the Native Hawaiian community. We are most fortunate to have the opportunity to conduct observations from this mountain.

The HARPS-N project was funded by the Prodex Program of the Swiss Space Office (SSO), the Harvard University Origin of Life Initiative (HUOLI), the Scottish Universities Physics Alliance (SUPA), the University of Geneva, the Smithsonian Astrophysical Observatory (SAO), the Italian National Astrophysical Institute (INAF), University of St. Andrews, Queen’s University Belfast, and University of Edinburgh.

This work has made use of data from the European Space Agency (ESA) mission {\it Gaia} (\url{https://www.cosmos.esa.int/gaia}), processed by the {\it Gaia} Data Processing and Analysis Consortium (DPAC, \url{https://www.cosmos.esa.int/web/gaia/dpac/consortium}). Funding for the DPAC has been provided by national institutions, in particular the institutions participating in the {\it Gaia} Multilateral Agreement.
This research has made use of the SIMBAD database and VizieR catalogue access tool, operated at CDS, Strasbourg, France. This research has made use of NASA's Astrophysics Data System. This research has made use of the NASA Exoplanet Archive, which is operated by the California Institute of Technology, under contract with the National Aeronautics and Space Administration under the Exoplanet Exploration Program.

This research has made use of the Exoplanet Follow-up Observation Program (ExoFOP; DOI: 10.26134/ExoFOP5) website, which is operated by the California Institute of Technology, under contract with the National Aeronautics and Space Administration under the Exoplanet Exploration Program.

This research is based on photographic data obtained using Oschin Schmidt Telescope on Palomar Mountain. The Palomar Observatory Sky Survey was funded by the National Geographic Society. The Oschin Shmidt Telescope is operated by the California Institue of Technology and Palomar Observatory. The plates were processed into the present compressed digital format with their permission. The Digitized Sky Survey was produced at the Space Telescope Science Institute (ST ScI) under US Goverment grant NAG W-2166.

L.D. thanks the Belgian Federal Science Policy Office (BELSPO) for the provision of financial support in the framework of the PRODEX Programme of the European Space Agency (ESA) under contract number 4000142531.

KR is grateful for support from UK's Science and Technology Facilities Council (STFC) via consolidated grant ST/V000594/1. 

DAT acknowledges the support of the Science and Technology Facilities Council (STFC).

The ULiege's contribution to SPECULOOS has received funding from the European Research Council under the European Union's Seventh Framework Programme (FP/2007-2013) (grant Agreement n$^\circ$ 336480/SPECULOOS), from the Balzan Prize and Francqui Foundations, from the Belgian Scientific Research Foundation (F.R.S.-FNRS; grant n$^\circ$ T.0109.20), from the University of Liege, and from the ARC grant for Concerted Research Actions financed by the Wallonia-Brussels Federation. MG is F.R.S-FNRS Research Director.

This research has made use of data obtained from or tools provided by the portal exoplanet.eu of The Extrasolar Planets Encyclopaedia.

\section*{Data Availability}

This paper includes raw data collected by the \textit{TESS} mission, which are publicly available from the Mikulski Archive for Space Telescopes (MAST, \url{https://archive.stsci.edu/tess}). Raw data collected by \textit{CHEOPS} can be found using the file keys in Table \ref{tab:CHEOPS_obs} at \url{https://cheops-archive.astro.unige.ch/archive_browser/}. Observations made with HARPS-N on the Telescopio Nazionale Galileo 3.6m telescope are available in Table \ref{tab:RV_HARPSN} in the Appendix Section \ref{sec:appendix_tables}. Observations made with TRES are available in Table \ref{tab:TRES_RVs} also in Appendix Section \ref{sec:appendix_tables}.



\bibliographystyle{mnras}
\bibliography{main} 



\section*{Affiliations}

$^{1}$University of Southern Queensland, Centre for Astrophysics, West Street, Toowoomba, QLD 4350 Australia\\
$^{2}$SUPA, Institute for Astronomy, University of Edinburgh, Blackford Hill, Edinburgh, EH9 3HJ, UK\\
$^{3}$Centre for Exoplanet Science, University of Edinburgh, Edinburgh, EH9 3HJ, UK\\
$^{4}$Mullard Space Science Laboratory, University College London, Holmbury St Mary, Dorking, Surrey, RH5 6NT, UK\\
$^{5}$School of Physics and Astronomy, University of Birmingham, Edgbaston, Birmingham B15 2TT, UK\\
$^{6}$Department of Physics, University of Warwick, Gibbet Hill Road, Coventry CV4 7AL, UK\\
$^{7}$Centre for Exoplanets and Habitability, University of Warwick, Coventry, CV4 7AL, UK\\
$^{8}$Fundaci\'on Galileo Galilei - INAF (Telescopio Nazionale Galileo), Rambla Jos\'e Ana Fern\'andez Perez 7, E-38712 Bre\~na Baja (La Palma), Canary Islands, Spain\\
$^{9}$Instituto de Astrof\'{\i}sica de Canarias, C/V\'{\i}a L\'actea s/n, E-38205 La Laguna (Tenerife), Canary Islands, Spain\\
$^{10}$Departamento de Astrof\'{\i}sica, Univ. de La Laguna, Av. del Astrof\'{\i}sico Francisco S\'anchez s/n, E-38205 La Laguna (Tenerife), Canary Islands, Spain\\
$^{11}$NASA Exoplanet Science Institute, IPAC, California Institute of Technology, Pasadena, CA 91125 USA\\
$^{12}$Department of Physics \& Astronomy, McMaster University, 1280 Main St W, Hamilton, ON, L8S 4L8, Canada\\
$^{13}$Astrobiology Research Unit, University of Liège, Allée du 6 août, 19, 4000 Liège (Sart-Tilman), Belgium\\
$^{14}$Space Sciences, Technologies and Astrophysics Research (STAR) Institute, Université de Liège, Allée du 6 Août 19C, 4000 Liège, Belgium\\
$^{15}$Institute of Astronomy, KU Leuven, Celestijnenlaan 200D, 3001 Leuven, Belgium\\
$^{16}$AIM, CEA, CNRS, Université Paris-Saclay, Université de Paris, F-91191 Gif-sur-Yvette, France\\
$^{17}$Department of Physics and Astronomy, The University of New Mexico, 210 Yale Blvd NE, Albuquerque, NM 87106, USA\\
$^{18}$NSF’s National Optical-Infrared Astronomy Research Laboratory, 950 N. Cherry Ave., Tucson, AZ 85719, USA\\
$^{19}$Astrobiology Research Unit, Université de Liège, Allée du 6 Août 19C, 4000 Liège, Belgium\\
$^{20}$Cavendish Laboratory, JJ Thomson Avenue, Cambridge CB3 0HE, UK\\
$^{21}$Department of Physics and Kavli Institute for Astrophysics and Space Research, Massachusetts Institute of Technology, Cambridge, MA 02139, USA\\
$^{22}$Juan Carlos Torres Fellow\\
$^{23}$Center for Astrophysics ${\rm \mid}$ Harvard {\rm \&} Smithsonian, 60 Garden St, Cambridge, MA 02138, USA\\
$^{24}$CAS Key Laboratory of Planetary Sciences, Purple Mountain Observatory, Chinese Academy of Sciences, Nanjing 210008, China\\
$^{25}$Proto-Logic Consulting LLC, Washington, DC\\
$^{26}$Department of Astrophysical and Planetary Sciences, University of 34 Colorado Boulder, Boulder, CO 80309, USA\\
$^{27}$Department of Earth, Atmospheric and Planetary Sciences, Massachusetts Institute of Technology, Cambridge, MA 02139, USA\\
$^{28}$Department of Aeronautics and Astronautics, MIT, 77 Massachusetts Avenue, Cambridge, MA 02139, USA

\appendix

\section{RV data}
\label{sec:appendix_tables}
\begin{landscape}
    \begin{table}
        \centering
        \caption{TRES radial velocity data and activity indicators.}
        \begin{tabular}{llllllllll}
             \hline\hline 
             BJD\textsubscript{UTC} & RV & $\sigma$\textsubscript{RV} & pha & Teff & $\sigma$\textsubscript{Teff} & Vrot & $\sigma$\textsubscript{Vrot} & CCF & SNRe \\
             (d) & (m s\textsuperscript{-1}) & (m s\textsuperscript{-1}) & & & & & & \\
             \hline
             \multicolumn{9}{l}{} \\
             2457674.500000 & 51286 & 21 & -71.41 & 0.0 & 0.0 & 0.0 & 0.5 & 0.931 & 20.1 \\
             2457920.500000 & 51308 & 21 & -61.76 & 0.0  & 0.0 & 0.8 & 0.5 & 0.940 & 23.5 \\
             2458654.500000 & 51279 & 21 & -33.00 & 0.0  & 0.0 & 1.9 & 0.5 & 0.915 & 16.2 \\
             2459468.500000 & 51256 & 21 & -1.11 & 0.0  & 0.0 & 1.5 & 0.5 & 0.947 & 23.3 \\
            \hline
        \end{tabular}
        \label{tab:TRES_RVs}
    \end{table}

    \begin{table}
        \centering
        \caption{HARPS-N radial velocity data and activity indicators.}
        \begin{tabular}{lllllllllllllll}
             \hline\hline 
             BJD\textsubscript{UTC} & RV (CCF) & $\sigma$\textsubscript{RV} (CCF) & RV (LBL) & $\sigma$\textsubscript{RV} (LBL) & BIS\textsubscript{span} & $\sigma$\textsubscript{BIS\textsubscript{span}} & FWHM & $\sigma$\textsubscript{FWHM} & H$\alpha$ & $\sigma$\textsubscript{H$\alpha$} & S-index & $\sigma$\textsubscript{S-index} & log$R'_{HK}$ & $\sigma$\textsubscript{log$R'_{HK}$}\\
             (d) & (m s\textsuperscript{-1}) & (m s\textsuperscript{-1}) & (m s\textsuperscript{-1}) & (m s\textsuperscript{-1}) & (m s\textsuperscript{-1}) & (m s\textsuperscript{-1}) & (km s\textsuperscript{-1}) & (km s\textsuperscript{-1}) &  & & &\\
             \hline
             \multicolumn{11}{l}{} \\
             2460165.657371 & 51336.09 & 2.48 & 51067.47 & 0.91 & 1932.44 & 4.96 & 2.65965 & 0.00496 & 0.639720 & 0.000441 & 0.803612 & 0.015745 & -5.706724 & 0.012155 \\
             2460166.649562 & 51337.54 & 6.53 & 51069.37 & 2.02 & 2133.89 & 13.06 & 2.67660 & 0.01306 & 0.624102 & 0.001419 & 0.624944 &    0.068003 & -5.872690 & 0.076935 \\
             2460171.600859 & 51338.04 & 3.02 & 51070.06 & 1.14 & 1949.39 & 6.04 & 
             2.65876 & 0.00604 & 0.672546 & 0.000691 & 0.891012	&   0.024193 & -5.644005 & 0.016166 \\
             2460172.701740 & 51330.58 & 2.36 & 51065.07 & 0.95 & 2001.89 & 4.72 & 
             2.66160 & 0.00472 & 0.640433 & 0.000522 & 0.844017 &   0.015760 & -5.676600 & 0.011352 \\
             2460173.618065 & 51335.21 & 2.21 & 51067.12 & 0.94 & 1885.06 & 4.42 & 2.65677 & 0.00441 & 0.663422 & 0.000570 & 0.841318 &   0.015271 & -5.678549 & 0.011049 \\
             2460174.667788 & 51334.79 & 2.24 & 51065.35 & 0.94 & 1914.42 & 4.47 & 2.66415 & 0.00447 & 0.649150 & 0.000555 & 0.899625 &   0.014965 & -5.638287 & 0.009869 \\
             2460193.607987 & 51338.87 & 2.42 & 51071.24 & 0.97 & 1914.82 & 4.85 & 
             2.66358 & 0.00485 & 0.665288 & 0.000552 & 0.886741 &   0.016841 & -5.646868 & 0.011328 \\
             2460196.589712 & 51337.73 & 2.64 & 51070.34 & 1.20 & 1958.99 & 5.27 & 2.66226 & 0.00527 & 0.869402 & 0.001124 & 1.526949 & 0.024313 & -5.347680 & 0.008211 \\
             2460198.605434 & 51338.94 & 3.10 & 51072.07 & 1.38 & 2006.43 & 6.21 & 2.66566 & 0.00621 & 0.682733 & 0.001251 & 0.780828 & 0.026873 & -5.724680 & 0.021622 \\
             2460199.632412 & 51339.12 & 1.89 & 51070.26 & 0.83 & 1957.92 & 3.77 & 2.66797 & 0.00377 & 0.625008 & 0.000478 & 0.859892 &   0.011799 & -5.665314 & 0.008281 \\
             2460208.578168 & 51338.38 & 2.97 & 51068.59 & 1.16 & 1926.13 & 5.93 & 2.66443 & 0.00593 & 0.635852 & 0.000746 & 0.889042 &  0.023664 & -5.645323 & 0.015860 \\
             2460215.542502 & 51341.32 & 3.86 & 51076.16 & 1.69 & 1812.75 & 7.73 & 2.65975 & 0.00773 & 0.631869 & 0.001897 & 0.714995 &  0.034402 & -5.781169 & 0.031525 \\
             2460218.560370 & 51339.23 & 3.73 & 51071.97 & 1.46 & 1956.52 & 7.46 & 2.65768 & 0.00746 & 0.602605 & 0.001217  & 0.662081 &  0.034328 & -5.832585 & 0.035411 \\
            \hline
        \end{tabular}
        \label{tab:RV_HARPSN}
    \end{table}
\end{landscape}

\section{Additional Figures}
\label{sec:add_figs}

\begin{figure*}
    \centering
    \includegraphics[width=0.45\linewidth]{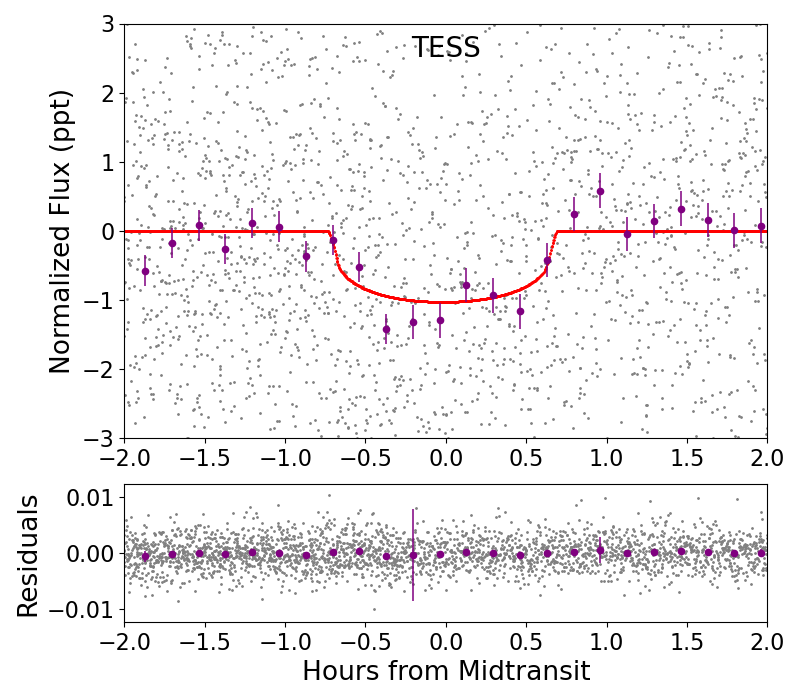}
    \includegraphics[width=0.45\linewidth]{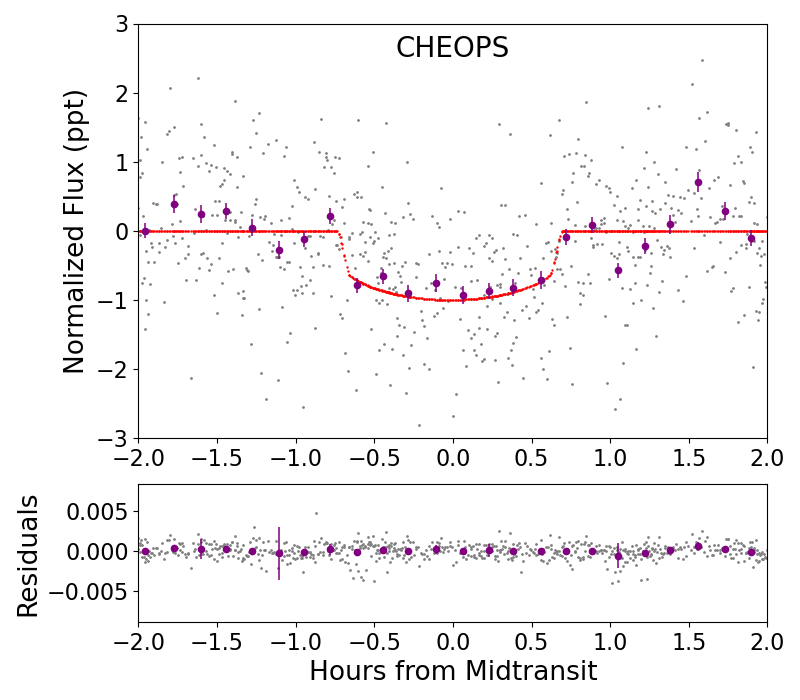}\\
    \includegraphics[width=0.33\linewidth]{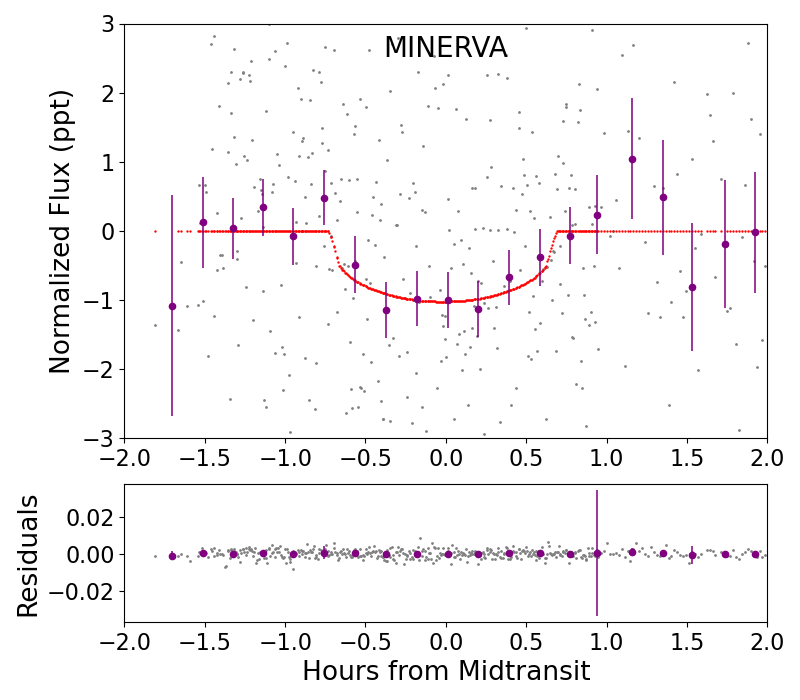}
    \includegraphics[width=0.33\linewidth]{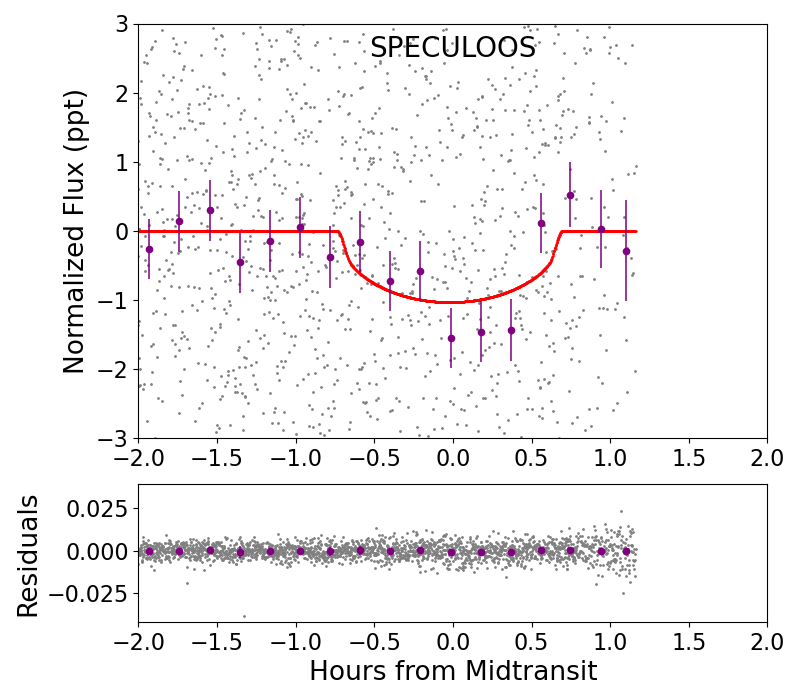}
    \includegraphics[width=0.33\linewidth]{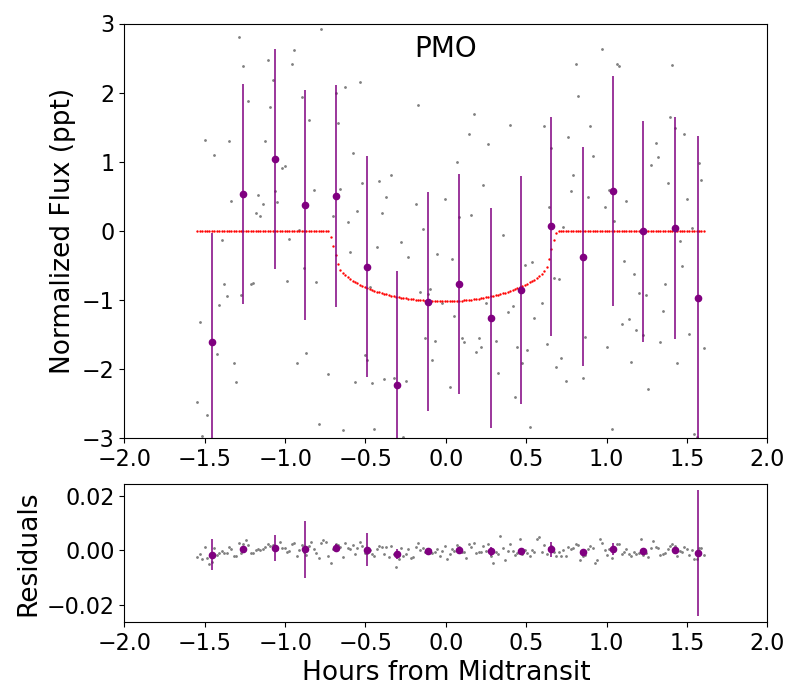}
    \caption{Phased transits of \thisstarb{} from fit results obtained with \texttt{dynesty}. Transit fit to the light curve data from \textit{TESS} (\textit{top left}), \textit{CHEOPS} (\textit{top right}), MINERVA (\textit{bottom left}), SPECULOOS (\textit{bottom}), and PMO (\textit{bottom right}). The short cadence (20s for \textit{TESS} sectors 42, 43, and 57, 120s for \textit{TESS} sector 70, 30s for \textit{CHEOPS}, 10s for all SPECULOOS telescopes, 30s for MINERVA T1, 60s for MINERVA T2, and 55s for PMO) fluxes are plotted in grey, the binned fluxes are overplotted in purple (10 minutes for space-based data and 11.5 minutes for ground-based data), and the fitted transit is shown by the red solid line.}
    \label{fig:dynesty_transits}
\end{figure*}


\bsp	
\label{lastpage}
\end{document}